\documentclass[pre,aps,showpacs,twocolumn,a4]{revtex4}

\usepackage{amsmath,amssymb}
\usepackage{graphicx}
\usepackage{psfrag}
\usepackage{array}
\usepackage{epsfig}
\renewcommand{\eqref}[1]{Eq.~(\ref{#1})}
\newcommand{\avg}[1]{\langle {#1} \rangle}
\newcommand{\pp}[2]{\frac{\partial #1}{\partial #2}}
\newcommand{\vect}{\mathbf}
\newcommand{\vabs}[1]{\left| #1\right|}
\newcommand{\dpath}{\mathcal{D}}
\newcommand{\order}{\mathcal{O}}
\newcommand{\lp}{\ell_\text{p}}
\newcommand{\lpar}{L_{\parallel}}
\newcommand{\lperp}{L_{\perp}}
\newcommand{\kT}{k_\text{B}T}
\newcommand{\rhotil} {\tilde{\rho}}

\newcommand {\etatil} {\tilde{\eta}}
\newcommand{\erfc}{{\mbox{erfc}}}
\newcommand{\erf}{{\mbox{erf}}}
\newcommand{\e}{\mbox{e}}
\renewcommand{\i}{\textrm{\mbox{i}}}

\begin{document}
\title{Entropic forces generated by grafted semiflexible polymers}

\author{Azam Gholami$^1$, Jan Wilhelm$^2$, Erwin Frey$^2$}

\affiliation{$^1$Hahn-Meitner-Institut, Abteilung Theorie,
  Glienicker Str. 100, D-14109 Berlin, Germany \\
  $^2$Arnold-Sommerfeld-Center for Theoretical Physics and CeNS, 
  Department of Physics, Ludwig-Maximilians-Universit\"{a}t M\"{u}nchen,\\
  Theresienstrasse 37, D-80333 M\"{u}nchen, Germany}


\date{\today}

\begin{abstract}
  The entropic force exerted by the Brownian fluctuations of a grafted
  semiflexible polymer upon a rigid smooth wall are calculated both
  analytically and by Monte Carlo simulations. Such forces are thought
  to play an important role for several cellular phenomena, in
  particular, the physics of actin-polymerization-driven cell motility
  and movement of bacteria like Listeria. In the stiff limit, where
  the persistence length of the polymer is larger than its contour
  length, we find that the entropic force shows scaling behavior. We
  identify the characteristic length scales and the explicit form of
  the scaling functions. In certain asymptotic regimes we give simple
  analytical expressions which describe the full results to a very
  high numerical accuracy.  Depending on the constraints imposed on
  the transverse fluctuations of the filament there are characteristic
  differences in the functional form of the entropic forces. In a
  two-dimensional geometry, the entropic force exhibits a marked peak.
\end{abstract}

\maketitle

\section{Introduction}

\begin{figure}[htbp]
  \begin{center}
    \input{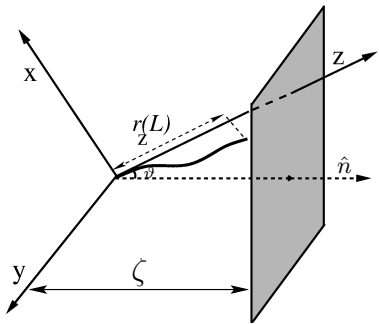}
    \caption{A smooth hard wall constrains the Brownian fluctuations
      of a grafted semiflexible polymer resulting in an entropic force
      on the  wall.}
    \label{fig:setup}
  \end{center}
\end{figure}

In a cellular environment soft objects like membranes and polymers are
subject to Brownian motion. As a result there are interactions between
them which are entropic in origin, i.e. a consequence of constraints
imposed on the Brownian fluctuations. For example, two parallel
membranes repel each other entropically with a potential that falls
off like a power law in the distance between them \cite{helfrich}.
Similarly, thermally fluctuating biopolymers like F-Actin and
microtubules may exert entropic forces on membranes or some other
obstacles; for an illustration see Fig.\ref{fig:setup}.
Though due to the same thermal fluctuations such forces have to be
distinguished from forces obtained by pulling on a biopolymer
\cite{force_extension,benetatos-frey:2004}. It will turn out that the
force-distance curves of these two cases have no resemblance at all in
a regime where thermal fluctuations play a role, which is generically
the case for all cytoskeletal filaments. Both types of forces are
thought to play a prominent role in cell motility and movement of
pathogens like listeria monocytogenes, that propel itself through the
cytoplasm of infected cells by constructing behind it a polymerized
tail of cross-linked actin filaments \cite{cykes}. Similarly, in a
crawling cell, the force generated from the growth of a collection of
actin fibers is responsible for the protrusion of cell membrane,
which are known as lamellipodia, filopodia, or microspikes according
to their shapes \cite{bray}. It seems that quite generally
polymerizing networks of actin filaments are capable of exerting
significant mechanical force, which are used by eukaryotic cells and
their prokaryotic pathogens to change shape or move. One type of force
is generated by fluctuating filaments at the leading edge of the
network. The length of the thermally fluctuating parts of these
polymers are typically $200\sim 300$nm, which is very short compared
to their persistence length $\lp\approx 17 \mu$m \cite{legoff}, such
that an analysis which considers these filaments as stiff seems
appropriate.

In this paper we will not enter into the debate on the particular
force generating mechanism responsible for all these different types
of cell motility, but rather give a detailed analysis of the entropic
forces which fluctuating stiff polymers exert on rigid walls. This
may serve as important input for future molecular models of
force generation in cellular systems.

Consider a semiflexible polymer with contour length $L$ and
persistence length $\lp$ with one end fixed both in position and
orientation to some rigid support (see Fig.\ref{fig:setup}), e.g.\/
the dense part of an actin gel. We choose coordinates such that the
grafted end is at the origin with the tangent fixed parallel to the
$z$-axis. Consider a rigid, smooth wall orthogonal to the
$x$-$z$-plane at a distance $\zeta$ from the origin. Let $\vartheta$
be the angle between the $z$-axis and the normal $\hat n$ of the wall.
If $\zeta$ is small enough, the wall will constrain the Brownian
fluctuations of the polymer leading to an increase in free energy with
respect to the unconstrained polymer. On time scales larger than the
equilibration time of the grafted polymer this results in an average
force $f$ exerted on the wall. Our goal is to calculate how the
entropic force $f$ depends on $\zeta$ and $\vartheta$ and the contour
length $L$ and the persistence length $\lp$ of the polymer.

We will proceed as follows. The following Section
\ref{sec:entropic_force} serves to introduce and discuss the various
types of thermodynamic forces which can be generated by fluctuating
semiflexible polymers. We will arrive at the conclusion that the
entropic forces discussed above are closely related to the probability
distribution of the free end of the clamped polymer. In Section
\ref{sec:orthogonal} we start our analysis of entropic forces with a
polymer grafted perpendicular to the wall. This chapter contains a
definition of the wormlike chain model and the basic idea of our
analytical calculations, which starting from the tip distribution
calculates the restricted free energy and the entropic force. The
analysis is complemented by Monte Carlo (MC) simulations, which both
show the range of validity of the analytical results and the crossover
from semiflexible to Gaussian chains. Details of the calculations are
deferred to the Appendices \ref{app:inverse_laplace},
\ref{app:saddle_point} and \ref{app:jacobi_transform_parallel}.
Section \ref{sec:oblique_angle} treats the technically more
complicated case of a polymer inclined at an angle $\vartheta$ with
respect to the wall. Here we obtain the entropic forces analytically
up to the numerical evaluation of some integrals. For some asymptotic
cases explicit analytical formula are again obtained. The MC
simulations in this chapter are restricted to a parameter range which
is close to the stiff limit, and mainly serve the purpose to define
the range of applicability of the anayltical results. Finally, in the
conclusion we give a discussion of our main results.

\section{Entropic forces and probability densities}
\label{sec:entropic_force}

According to the {\em wormlike chain model} \cite{kratky-porod:49,
  saito-takahashqi-yunoki:67}, the elastic energy of a given
configuration $\vect r(s)$, parameterized in terms of the arc length
$s \in [0,L]$, is given by
\begin{eqnarray}
  \label{eqn:wormlike_chain}
  \beta H = \frac{\lp}{2} \int_0^L d s
            \left(\frac{\partial\vect t(s)}{\partial s}\right)^2 \, .
\end{eqnarray}
Here $\vect t(s) = \partial \vect r(s)/\partial s \equiv \dot {\vect
  r} (s)$ is the local tangent to the contour $\vect r(s)$ and $\lp =
\kappa/\kT$ is the persistence length with $\kappa$ the polymer's
bending modulus, and $\beta = 1/\kT$. As the polymer is considered to
be inextensible, we have $\vabs{\vect t(s)} = 1$ for all $s$, i.e. the
tangent vectors are restricted to the unit sphere.

In a cellular environment biopolymers are flexed by Brownian motion,
i.e. they exhibit thermal fluctuations in their shape. This mere fact
makes for a rich mechanic response genuinly different from its
classical analogue, a rigid beam. Consider a polymer whose position
(not its orientation) is fixed at one end and one is pulling on its
other end, a typical situation encountered in an experiment using
optical or magnetic tweezers. Then there is no unique force-distance
relation.  It actually matters whether one pulls at constant force $f$
and measures the resulting average distance $\avg{r} (f)$ or vice
versa.  Results for the constant force ensemble are thoroughly
discussed in Ref.\cite{force_extension}.  In a constant distance
ensemble the probability density distribution of the end-to-end
distance $P(r)$ provides the necessary information
\cite{wilhelm-frey:96}. It defines a free energy $F(r)=-\kT \ln P(r)$
from which the average force may be derived by differentiation,
$\avg{f}(r)=-\partial F(r) / \partial r$ \cite{fkws_review:98}.

Here we are interested in the force a fluctuating filament exerts on a
rigid obstacle which is fixed in its position \cite{footnote:1}. The
polymer's end facing the obstacle is considered as free to fluctuate
and only its proximal end is fixed in position and orientation; see
Fig.\ref{fig:setup}. Since there are no direct forces between polymer
and obstacle the force exerted on the wall is solely due to the steric
constraints imposed on the filament. This suggests to use the term
``entropic forces'', frequently used in analogous physical situations
\cite{lubensky:review}. However, this should not leave the reader with
the wrong impression that there are different physical origins for
entropic forces and those discussed in the preceeding paragraph. It is
merely the type of ``boundary condition'' imposed on the thermal
fluctuations which leads to their (drastically) different character.

For getting acquainted with the problem let us consider the simplest
case, a grafted polymer whose one end and tangent is fixed such that
it is oriented perpendicular to a smooth wall (Fig.\ref{fig:setup}
with $\vartheta = 0$). The presence of the wall allows only for those
polymer configurations which are entirely in the halfspace to the left
of the wall.  Since we are mostly interested in stiff polymers (which
have a low probability for back-turns) this restriction may be
approximated as a constraint solely on the position of the polymer tip
facing the wall, $r_z (L) \leq \zeta$; later in Section
\ref{sec:entropic_forces_explicit} we will show some simulation data
going beyond this approximation.

To derive the average force acting on the wall we consider a wall
potential $U (\zeta - r_z (L))$ for the free polymer tip, which at the
end of the calculation will be reduced to a hard wall potential. For
now picture a steep potential which rises rapidly for $r_z (L) \to
\zeta$. Then, the ensemble average for the force the polymer tip
excerts perpendicular to the wall reads
\begin{eqnarray}
  \avg{f_\parallel} (\zeta)
  = \frac{1}{{\cal Z}_\parallel (\zeta)}  \,
    \int \dpath [\vect r(s)] \, 
    \e^{- \beta (H+U)} \, 
    \frac{\partial U}{\partial r_z (L)} \, .
\label{eq:stat_mech_def_ef}
\end{eqnarray}
Here the partition sum 
\begin{eqnarray}
    {\cal Z}_\parallel (\zeta)
    = \int \dpath [\vect r(s)] \, 
    \e^{- \beta (H+U)} \, 
\end{eqnarray}
is a path integral over all polymer configurations compatible with the
boundary conditions imposed on the distal and free end of the grafted
polymer, where the measure is taken such that the partition sum
without a constraining wall ($U=0$) is normalized to $1$.  This is now
a thermodynamic force. In an actual experiments it is obtained by a
time average with an averaging time much larger than the equilibration
time for the grafted polymer. This force would also be measured in
an experiment where a large number of independent and identical
polymers push against the same wall.   

Since the wall potential depends only on the difference between the
position of the polymer tip and the wall we may rewrite the entropic
force in \eqref{eq:stat_mech_def_ef} as
\begin{eqnarray}
    \avg{f_\parallel} (\zeta)
  = \kT \frac{\partial}{\partial \zeta} 
    \ln {\cal Z}_\parallel (\zeta) \, .
\label{eq:entropic_force_lnZ}
\end{eqnarray}
Upon definig a free energy of the confined polymer as
\begin{eqnarray}
  \cal{F}_\parallel(\zeta)
  &=& -\kT \ln {\cal Z}_\parallel(\zeta) \; ,
\end{eqnarray}
the entropic forces again reads as a spatial derivative of a free
energy
\begin{eqnarray}
  \avg{f_\parallel} (\zeta)
  = - \frac{\partial}{\partial \zeta} \cal{F}_\parallel(\zeta) \, .
\end{eqnarray}
The physical interpretation of this free energy becomes clear as one
goes to the hard wall limit. Then, the partition function reduces to
\begin{eqnarray}
   {\cal Z}_\parallel(\zeta) 
   &=& \int \dpath [\vect r(s)]  \, 
       \Theta(\zeta - r_z(L)) \, \e^{-\beta H}
   \nonumber \\
   &=:& \avg{\Theta(\zeta - r_z(L))}_0 \; ,
\end{eqnarray}
where the subscript $0$ indicates that the average is now taken with
respect to the bending Hamiltonian only. The $\Theta$-function,
defined such that $\Theta (x) = 1$ for $x > 0$ and zero elsewhere,
indicates that only those configurations are counted with the position
of the polymer tip to the left of the wall. Hence, as for the fixed
distance ensemble in a pulling experiment, the free energy results
from a quantity measuring the number of configurations obeying the
imposed constraint, where each configuration is weighted by a
Boltzmann factor for the bending energy.

It is useful to rewrite the partition function as
\begin{eqnarray}
  {\cal Z}_\parallel(\zeta)
  &=& \int_{-L}^L dz \, \Theta(\zeta - z) \avg{\delta(z - r_z(L))}_0
  \nonumber \\
  &=& \int_{-L}^\zeta dz \,  P_\parallel(z) \; ,
\end{eqnarray}
where $ P_\parallel(z) = \avg{\delta(z - r_z(L))}_0 $ is the
probability density to find the $z$-coordinate of the polymer's free
end at $z$, irrespective of its transverse coordinates. It identifies
the restricted partition sum as the cummulative distribution function
corresponding to the probability density $P_\parallel (z)$. One may
then write the entropic force in the alternative form
\begin{eqnarray}
  \avg{f_\parallel} (\zeta)
  = \kT \, \frac{P_\parallel (\zeta)}{{\cal Z}_\parallel(\zeta)} \, . 
\label{eq:entropic_force-best}
\end{eqnarray}
Upon multiplying this formula by $d \zeta$ it may be interpreted as
follows. The work done on the wall upon displacing it by an
infinitesimal distance $d\zeta$ equals the thermal energy scale $\kT$
times a conditional probability $P_\text{left} (\zeta) d\zeta =
P_\parallel (\zeta) d\zeta/ {\cal Z}_\parallel(\zeta)$, which measures
the probability that the position of the polymer tip is within a
distance $d\zeta$ from the wall given that the polymer is in the left
halfspace.

Since the probability density for the position of the polymer tip
$P(\vect x,z)$ is actually a function of the position perpendicular
and transverse to the wall, \eqref{eq:entropic_force-best} immediately
suggests that one could define a {\em local} entropic pressure.
Indeed, upon generalizing the above arguments one may write
\begin{eqnarray}
  p (\vect x,\zeta)
  &=&\frac{-1}{{\cal Z}_\parallel (\zeta)} 
    \int \! \dpath [\vect r] \, 
    \frac{\partial U}{\partial \zeta} 
    \delta(\vect x \!-\! \vect r_\perp (L)) \, 
    \e^{- \beta (H+U)} 
 \nonumber \\
 &=& \frac{\kT}{{\cal Z}_\parallel (\zeta)} \, 
     \frac{\partial}{\partial \zeta}
     \avg{\Theta(\zeta - r_z(L)) \, \delta(\vect x - \vect r_\perp (L))}_0
 \nonumber \\
 &=& \kT \; \frac{P(\vect x, \zeta)}{{\cal Z}_\parallel (\zeta)}
\label{eq:stat_mech_def_ef2}
\end{eqnarray}
for the entropic pressure, i.e. the force per unit area exerted
locally at $\vect x$ on the wall. Again, the entropic force is given
by the thermal energy scale times a conditional probability density,
which now measures the probability of finding the polymer tip at a
particular site $\vect x$ on the wall conditioned on the polymer
configuration being to the left of the wall. Pictorially, one may say
that the local pressure is given by $\kT$ times the number of
``collisions'' of the polymer with the wall per unit area, a reasoning
which is frequently used in scaling analyses. 

The total force is, of course, obtained by integrating over this local
pressure, $\avg{f_\parallel} (\zeta) = \int d \vect x \, p(\vect
x,\zeta)$. In addition, one may now also define an entropic torque as
has recently been done for a rigid rod facing a planar wall
\cite{roth:entropic_torque}; we leave this issue for future investigations.

Generalizing the above ideas  suggests
to introduce an effective local free energy per unit area as 
\begin{eqnarray}
  {\cal F} (\vect x, \zeta) 
  = - \kT  \; \int^\zeta d z  
      \frac{P(\vect x, \zeta)}
           {{\cal Z}_\parallel (\zeta)} \, , 
\end{eqnarray}
which is useful in applications where the obstacle is actually not
rigid but soft with some internal elasticity, e.g. a membrane, whose
dynamics is much slower than the equilibration time of the
polymer. Then the elastic energy describing membrane bending and the above  
effective free energy may just be added to describe the combined
system. Of course, such a description fails if time scales for the
dynamics of both soft objects are comparable.

Our main conclusion in this section is that entropic forces generated
by a grafted stiff polymer can be reduced to the calculation of the
probability distribution of the polymer tip. For a polymer constrained
to two dimensions this distribution function has been found to show
quite interesting behavior such as bimodality in the transverse
displacement of the free end \cite{lattanzi-munk-frey:2004}.  This
pronounced feature of the distribution function has recently been
rationalized upon exploiting an interesting analogy to a random walker
in shear flow \cite{benetatos-munk-frey:2005}.

\section{Polymer orthogonal to a wall}
\label{sec:orthogonal}

In this section we are going to calculate the entropic force generated
by a grafted polymer whose orientation is on average perpendicular to
the wall. It illustrates the basic idea of our analytical calculations
for the simplest geometry.

\subsection{Weakly bending limit: mode analysis}

In evaluating the distribution function analytically we restrict
ourselves to the limit of a {\em weakly bending filament}. In other
words, we consider the persistence length $\lp$ to be large enough
compared to the total contour length $L$, such that the statistical
weight of configurations with small sharp bends will be negligible.
The key small dimensionless quantity will be the {\em stiffness
  parameter}
\begin{eqnarray}
  \label{eq:stiffness_parameter}
  \varepsilon = L/\lp
\end{eqnarray}
and we will refer to the weakly bending limit also as the {\em stiff
  limit}.

For small $\varepsilon$, the transverse components, $t_x(s)$ and
$t_y(s)$, of the tangent vector $\vect t (s)$ will be small for all
$s$.  While the condition $\vabs{\vect t(s)} = 1$ would suggest a
parameterization of $\vect t(s)$ in terms of polar coordinates or
Euler angles, for reasons that will become apparent later we want to
have independent variables that appear in a symmetric way in the
integrand. Thus we choose to parameterize $\vect t$ by
\begin{eqnarray}
  \label{eqn:tangent_approx}
  \vect t = \frac{1}{\sqrt{1+a_x^2 + a_y^2}}
   \left(\begin{array}{c} a_x \\ a_y \\ 1 \end{array}\right),
\end{eqnarray}
where we dropped all arguments $s$ for brevity; the generalization to
$d$ spatial dimensions is obvious.

The boundary conditions at the ends of the polymer are
\begin{subequations}
\begin{eqnarray}
\vect t(0) &=& (0,0,1)^T \quad \text{(clamped end)} \, , \\
\dot{\vect t}(L) &=& (0,0,0)^T \quad \text{(free end)} \, .
\end{eqnarray}  
\label{eq:bc}
\end{subequations}
This translates into $\vect a (0) = (a_x(0),a_y(0))^T = (0,0)^T$ and
$\dot {\vect a} (L) = (\dot a_x(L), \dot a_y(L))^T = (0,0)^T$. We thus
can choose a Fourier representation, or in other words a {\em normal mode
decomposition}
\begin{eqnarray}
  \label{eqn:fourier_tangent}
  a_x(s) = \sum_{k=1}^\infty a_{x,k}
           \sin \left( \lambda_k \frac{s}{L} \right)
\end{eqnarray}
with eigenvalues
\begin{eqnarray}
   \lambda_k = \frac{\pi}{2} \, (2k-1) \, ,
\end{eqnarray}
and Fourier (normal mode) amplitudes 
\begin{eqnarray}
  \label{eqn:inverse_fourier}
  a_{x, k} = \frac{2}{L} \int_0^L d  s \, a_x(s)
  \sin \left( \lambda_k \frac{s}{L} \right) \, ,
\end{eqnarray}
and similar for $a_y(s)$. To second order in the Fourier amplitudes
the location of the end-point along the $z$-axis reads
\begin{eqnarray}
  \label{eqn:rz_by_a}
  r_z(L) &=& \int_0^L d  s \, t_z(s)
  \nonumber \\
  &\approx& L - \frac{1}{2} \int_0^L d  s \,
   \left[ a_x^2(s) + a_y^2(s) \right]
  \nonumber \\
  &=& L - \frac{L}{4}
  \sum_{k=1}^\infty \left[ a_{x, k}^2 + a_{y, k}^2 \right] \,.
\end{eqnarray}
Similarly, we find for the Hamiltonian to second order
\begin{eqnarray}
  \label{eqn:hamil_harm}
  \beta H 
  \approx \frac{\lp}{4 L}
  \sum_{k=1}^\infty \lambda_k^2
      \bigl[ a_{x, k}^2 + a_{y, k}^2 \bigr] \,.
\end{eqnarray}

\subsection{Moment generating function}
\label{sec:mgf}

For calculating the probability density function $P_\parallel(z)$ we
follow a procedure outlined in Ref.\cite{wilhelm-frey:96} and
consider the moment generating function
\begin{eqnarray}
  \label{eqn:def_mogenfun}
  {\cal P}_\parallel (f) &:=& \avg{\e^{-f (L-r_z(L))}}_0
  \nonumber \\
  &=& \int_{-L}^{L} dz \, \e^{-f (L-z)} P_\parallel (z)
  \nonumber \\
  &=& \int_0^{2L} d \rho \, \e^{-f \rho} P_\parallel (L-\rho) \, .
\end{eqnarray}
Note that thermal averages have to be evaluated using the bare elastic
free energy, \eqref{eqn:wormlike_chain}.  Since for stiff chains
configurations with large values for the stored length
(``compression'') $\rho = L-z$ are rather unlikely, we can extend the
upper boundary of the integral in the last line of the preceding
equation to infinity. This allows us to write the moment generating
function as the Laplace transform of the distribution function
$P_\parallel (z)$
\begin{eqnarray}
   \label{eqn:laplace_transform_parallel}
   {\cal P}_\parallel (f)
   = \int_0^\infty d \rho \, \e^{-f \rho} P_\parallel (L-\rho) \, .
\end{eqnarray}
For $f=0$ the latter equation reduces to the normalization condition
of the probability density function $P_\parallel (z)$ such that ${\cal
  P}_\parallel (0)=1$.

Combining Eqs.~\ref{eqn:rz_by_a}, \ref{eqn:hamil_harm} and
\ref{eqn:def_mogenfun} the moment generating function can be put into
the following path integral form
\begin{eqnarray}
 {\cal P}_\parallel (f) = \int 
  \dpath [{\vect a} (s)]
  \exp
  \left\{
    - \frac12 \int_0^L \! ds
      \left[ \lp \dot{\vect a}^2
             + f  \vect a^2 \right]
  \right\} 
\end{eqnarray}
with the boundary conditions given by \eqref{eq:bc}.  This path
integral is easily evaluated upon using the Fourier representation of
the transverse tangent fields, Eq.  (\ref{eqn:fourier_tangent}), and
noting that to harmonic order fluctuations in all transverse
directions are statistically independent. We find in $d$ spatial
dimensions 
\begin{eqnarray}
  \label{eqn:mgf_result}
  {\cal P}_\parallel (f)
  &=& \left(
    \int \prod_{k=1}^\infty 
    \frac{d a_k}{\cal N}
    \exp \left\{ -\frac14
                  \left[
                  \frac{\lambda_k^2 \lp}{L}
                  + f L
                  \right] a_k^2
                  \right\}
    \right)^{(d-1)}
  \nonumber \\
  &=& \prod_{k=1}^\infty
      \left( 1 + \frac{fL^2}{\lp \lambda_k^2} \right)^{-(d-1)/2} \, ,
\end{eqnarray}
where the normalization factor ${\cal N}$ of the path integral was
chosen such that ${\cal P}_\parallel (0)=1$.  If $f\in \mathbb R_+$
the product may be rewritten as \cite{Hansen}
\begin{eqnarray}
  \label{eqn:mgf_result2}
  {\cal P}_\parallel (f)
= \left( \cosh \sqrt{\frac{fL^2}{\lp}} \right)^{-\frac12(d-1)} \, .
\end{eqnarray}
Note that the moment generating function, which also depends on the
length scales $L$ and $\lp$, has the scaling form
\begin{eqnarray}
   {\cal P}_\parallel (f,L,\lp)
   =   \tilde {\cal P}_\parallel (f \lpar) \, ,
\end{eqnarray}
where we have defined the characteristic {\em longitudinal length
  scale}
\begin{eqnarray}
  \lpar := \frac{L^2}{\lp} \, .
\end{eqnarray}
The formulas \eqref{eqn:mgf_result} and \eqref{eqn:mgf_result2} are
the basis for all subsequent calculations in this section, which are
basically different forms of performing the inverse Laplace transform.

For future reference and comparison with the entropic forces we close
this subsection with a discussion of the force-extension relation in the
fixed force ensemble. It simply follows as the first moment of the
moment generating function
\begin{eqnarray}
  \avg{r_z(L)}_f 
  &=&  L + \pp{\ln {\cal P}_\parallel (f)}{f}
  \nonumber \\
  &=&L\left( 1 - \frac{L(d-1)}{4\lp}
                 \frac{\tanh \sqrt{f\lpar}}
                      {\sqrt{f\lpar}}
      \right) \, ,
\end{eqnarray}
where $f$ is the external force in units of the thermal energy $\kT$.
In the limit of small external forces this reduces to
\begin{eqnarray}
   \avg{r_z(L)}_f =
   L \left[
     1 - \frac{d-1}4 \, \frac{L}{\lp}
       + \frac{d-1}{12}\,
         \left( \frac{L}{\lp} \right)^2
         f L \right] \, ,
\end{eqnarray}
which identifies $\lpar$ as $4/(d-1)$ times the equilibrium stored
length due to thermal fluctuations. We also recover the effective
linear spring coefficient $k_\parallel = 12\kappa^2/(d-1)\kT
L^4$, which was previously calculated in Ref.\cite{kroy-frey:96}. For
strong stretching forces the extension saturates asymptotically as
\begin{eqnarray}
    \avg{r_z(L)}_f =
   L \left[ 1 - \frac{L~(d-1)}{4~\lp\sqrt{f\lpar}} \right] \, .
\end{eqnarray}
In the limit of large compressional forces the weakly bending rod
approximation breaks down and one has to use different approaches to
evaluate the force-extension relation \cite{fkws_review:98}. 

\subsection{Probability density for the position of the polymer tip:
  analytical and MC results in 3d}
\label{sec:P_parallel}

\begin{figure}[htbp]
  \begin{center}
    \psfrag{x}{\hspace{-1cm}$\rhotil=(L-z)/L_{\parallel}$}
    \psfrag{y}{\hspace{-0.3cm}$\tilde P_\parallel(\rhotil)$}
    \rotatebox{-90}{\includegraphics[height=\columnwidth]
      {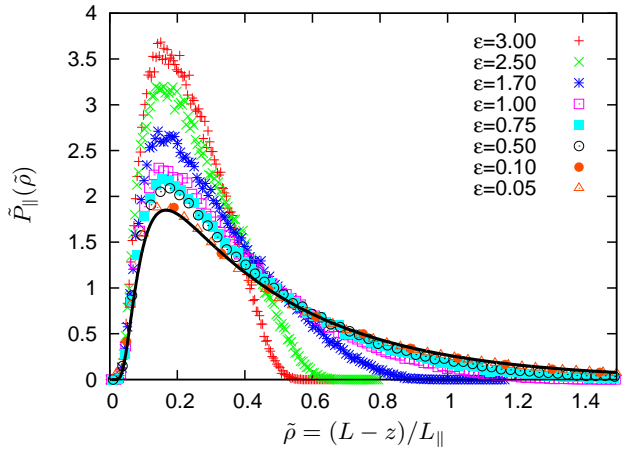}}
    \caption{Scaling function $\tilde P_\parallel(\rhotil)$ 
      (solid line) in 3d for the probability density to find the free
      end of a grafted semiflexible polymer in a plane defined by
      $r_z(L) = z$ or equivalently with a reduced stored length
      $\rhotil$. For comparison MC data are given for a series of
      stiffness parameters $\varepsilon = L / \lp$ indicated in the
      graph.  Deviations from the scaling curve in the stiff limit
      become significant for $\varepsilon = 0.5$ and larger.}
    \label{fig:P_parallel}
  \end{center}
\end{figure}

We now return to the distribution function and the resulting entropic forces.
Upon performing the inverse Laplace transform one gets (for
details of the calculations see Appendix \ref{app:inverse_laplace3d})
\begin{eqnarray}
  \label{eqn:P_form1}
  P_\parallel(z) &=& \frac{2}{\lpar} \sum_{k=1}^\infty (-1)^{k+1}
  \lambda_k
  \exp \left[- \lambda_k^2 \frac{L-z}{\lpar} \right] \, .
\end{eqnarray}
Inspection of Eq. (\ref{eqn:P_form1}) immediately tells us that it can 
be written in scaling form
\begin{eqnarray}
P_\parallel (z,L,\lp) = \lpar^{-1} \tilde P_\parallel (\rhotil) \, ,
\end{eqnarray}
where we have made the dependence of the probability density
on $L$ and $\lp$ explicit and introduced the scaling
variable
\begin{eqnarray}
\rhotil = \frac{L - z}{\lpar}
\end{eqnarray}
measuring the compression of the filament in units of $\lpar$. This
implies that data for the probability density of the polymer tip can
be rescaled to fall on a {\em scaling function} $\tilde P_\parallel
(\rhotil)$, shown as the solid curve in Fig.\ref{fig:P_parallel}. Of
course, since the analytical calculations are based on the mode
analysis in the weakly bending limit such a universal scaling curve is
obtained only for small enough stiffness parameters $\varepsilon$.

The probability density is strongly peaked towards full stretching,
$\rhotil \rightarrow 0$, and falls off exponentially for large
$\rhotil$, such that for  $\rhotil \geq 0.3$  
\begin{eqnarray}
  \tilde P^>_\parallel (\rhotil) 
  = \pi \exp \left( - \frac 14 \pi^2 \rhotil \right)
\label{eq:app_tip_distribution_3d_greater}
\end{eqnarray}
is already an excellent approximation. The series expansion given by
\eqref{eqn:P_form1} converges well for all values of $z$ well below
$L$, but its convergence properties become increasingly worse if $z$
approaches $L$. As detailed in Appendix \ref{app:inverse_laplace2d}
one may also derive an alternative series representation of the tip
distribution function which converges well close to full stretching
\begin{eqnarray}
  \label{eqn:P_form2}
   \tilde P_\parallel(\rhotil) = 
    \sum_{l=0}^\infty (-1)^{l} \;
    \frac{2l+1}{\sqrt{\pi \rhotil^3}} 
    \exp \left[- \frac{(l+\frac12)^2}{\rhotil} \right] \, .
\end{eqnarray}
Already the first term of \eqref{eqn:P_form2} 
\begin{eqnarray}
  \tilde P_\parallel^< (\rhotil)
   = \frac{1}{\sqrt{\pi {\rhotil}^3}} \,
     \exp \left( -\frac{1}{4\rhotil} \right)  \, ,
\label{eq:app_tip_distribution_3d_smaller}
\end{eqnarray}
gives an excellent fit for $\rhotil \leq 0.3$. In particular, it
captures the main feature of the distribution function, namely its
maximum close to full stretching. The same approximate expression may
also be obtained by evaluating the inverse Laplace transform using the
method of steepest descent; see Appendix \ref{app:saddle_point}. The
asymptotic results given in \eqref{eq:app_tip_distribution_3d_smaller}
and \eqref{eq:app_tip_distribution_3d_greater} taken together give a
representation of the scaling curve to a very high numerical accuracy.
They are the analogues of the results found in
Ref.\cite{wilhelm-frey:96} for a freely fluctuating filament; see also
Ref.\cite{fkws_review:98}.

The MC data shown in Fig.\ref{fig:P_parallel} have been obtained by
using a standard algorithm for a discretized wormlike chain, similar
to the one described in Ref.\cite{wilhelm-frey:96}. As expected, the
MC results agree very well with the analytical calculations for small
values of $\varepsilon$. From Fig.\ref{fig:P_parallel} we can read off
that the asymptotic stiff scaling regime remains valid up to
stiffness parameters $\varepsilon \approx 0.1$; even for $\varepsilon
= 0.5$ the shape of the scaling function resembles the MC data quite
closely. As the polymer becomes more flexible the shape asymptotically
becomes Gaussian; for $\varepsilon = 3$ a skew is still noticeable.
Note that in the parameter range given in Fig.\ref{fig:P_parallel} the
width of the rescaled probability densities stays approximately the
same and is hence well characterized by the longitudinal scale
$\lpar$.

\subsection{Confinement free energy and entropic forces: 3d}
\label{sec:force_parallel_series}

Now we are in a position to calculate the restricted partition sum
(cummulative probability distribution) ${\cal Z}_\parallel(\zeta) =
\int_{-L}^\zeta d z P_\parallel(z)$ by (formally) integrating the
series expansion \eqref{eqn:P_form1} term by term. This gives
\begin{eqnarray}
  \label{eqn:Z_form1}
  {\cal Z}_\parallel(\zeta)
  &=& 1 - \int_\zeta^L dz \, P_\parallel (z)
  \nonumber \\
  &=& 1 - 2 \sum_{k=1}^\infty (-1)^{k+1} \lambda_k^{-1} 
      \left(1 - \e^{-\lambda_k^2 (L-\zeta)/\lpar}\right)
  \nonumber \\
  &=& 2 \sum_{k=1}^\infty (-1)^{k+1} \lambda_k^{-1}
      \e^{-\lambda_k^2 (L-\zeta)/\lpar} \, ,
\end{eqnarray}
where in the first line we used the normalization of $P_\parallel (z)$
and in the final line the identity \cite{Stegun}
\begin{eqnarray}
  \sum_{k=1}^\infty (-1)^{k+1} \frac{1}{2k-1} =
    \frac{\pi}{4} \,.
\end{eqnarray}
The series expansion in \eqref{eqn:Z_form1} converges well for all values
of $\zeta$ well below $L$. Alternatively, one may start from
\eqref{eqn:P_form2} and derive
\begin{eqnarray}
  \label{eqn:Z_parallel_smallrho}
  {\cal Z}_\parallel(\zeta)
  &=& 1+2\sum_{k=1}^\infty (-1)^k
    \erfc \left( \frac{\lambda_k/\pi}{\sqrt{(L-\zeta)/\lpar}} \right) \, ,
\end{eqnarray}
which is well behaved for $\zeta$ close to $L$, and dominated by its
first term.  A second method to obtain
\eqref{eqn:Z_parallel_smallrho} can be found in Appendix
\ref{app:jacobi_transform_parallel}.

\begin{figure}[htbp]
  \begin{center}
    \psfrag{x}{\hspace{-1cm}$\etatil=(L-\zeta)/L_{\parallel}$}
    \psfrag{f}{\hspace{-0.3cm}$\tilde F_\parallel(\etatil)$}
    \rotatebox{-90}{\includegraphics[height=\columnwidth]
      {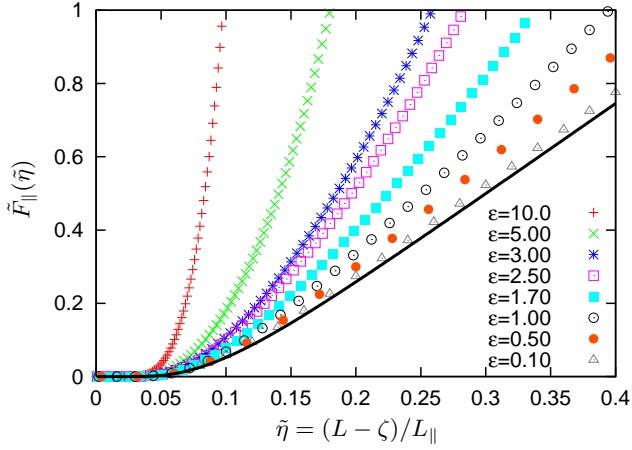}}
    \caption{Confinement free energy $\tilde{\cal F}_\parallel (\etatil)$ 
      of a grafted polymer constrained by a rigid wall in 3d as a
      function of the reduced minimal stored length $\etatil =
      (L-\zeta)/\lpar$. The solid line gives the scaling function
      obtained in the limit of a weakly bending rod. Symbols represent
      MC data for different values of the stiffness parameter
      $\varepsilon$ as indicated in the graph.}
    \label{fig:FreeEnergy__universal}
  \end{center}
\end{figure}

>From both series expansions it is evident that the restricted
partition sum has the scaling property
\begin{eqnarray}
 {\cal Z}_\parallel(\zeta, L, \lp) = \tilde {\cal Z}_\parallel (\etatil) \, ,
\end{eqnarray}
where we have introduced the scaling variable 
\begin{eqnarray}
\etatil = \frac{L-\zeta}{\lpar}\, , 
\end{eqnarray}
which measures the minimal stored length (compression) $\eta =
L-\zeta$ of the filament in units of $\lpar$.  The confinement free
energy, $\tilde {\cal F}_\parallel (\etatil) = - \kT \ln \tilde {\cal
  Z}_\parallel (\etatil)$, corresponding to this partition function is
shown in Fig.\ref{fig:FreeEnergy__universal}.
Again, the universal scaling function describes the MC data well for
$\varepsilon \leq 0.1$. Note that for all values of $\etatil$ and the
stiffness parameter $\varepsilon$ the free energy is convex. This will
turn out to be an important feature which distinguishes the 3d and 2d case.

Upon using \eqref{eq:entropic_force-best} for the entropic force we
find
\begin{eqnarray}
  f_\parallel (\zeta)
  = \frac{\kT}{\lpar} 
      \frac{\tilde P_\parallel(\etatil)}
           {\tilde {\cal Z}_\parallel(\etatil)} 
\label{eq:entropic_force-best:3d}
\end{eqnarray}
which immediately shows its scaling behavior and identifies $\kT /
\lpar$ as the characteristic force scale. It is up to a prefactor
identical to the critical force
\begin{eqnarray}
 f_c = \frac{\pi^2 \kappa}{4L^2} = \frac{\pi^2}{4} \, \frac{\kT}{\lpar}
\end{eqnarray}
for the buckling instability of a classical Euler-Bernoulli beam
\cite{landau-lifshitz-7}. It suggest to rewrite the entropic force as
\begin{eqnarray}
  \label{eqn:fparallel}
  f_\parallel(\zeta,L,\lp) 
  = f_c \, \tilde f_\parallel(\etatil) \,,
\end{eqnarray}
with the scaling function 
\begin{eqnarray}
  \tilde f_\parallel(\etatil) :=
  \frac{4}{\pi^2}\frac{\tilde P_\parallel(\etatil)}
  {\tilde {\cal Z}_\parallel(\etatil)} \, .
\end{eqnarray}
\begin{figure}[htbp]
  \begin{center}
    \psfrag{x}{\hspace{-1cm}$\etatil=(L-\zeta)/L_{\parallel}$}
    \psfrag{f}{\hspace{-0.3cm}$\tilde f_\parallel(\etatil)$}
    \rotatebox{-90}{\includegraphics[height=\columnwidth]
      {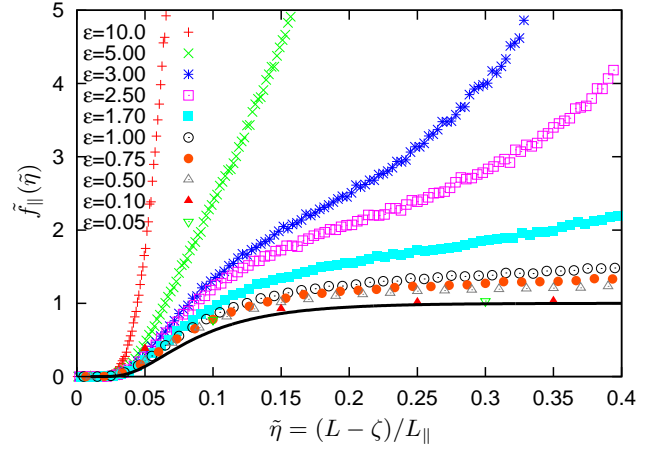}}
    \caption{Scaling function $\tilde f_\parallel(\etatil)$ for the 
      entropic force exerted on a wall at a distance $\zeta$ from the
      grafted end as a function of the scaling variable $\etatil =
      (L-\zeta)/\lpar$. MC data for different stiffness parameters
      $\varepsilon$ as indicated in the graph.}
    \label{fig:f_universal}
  \end{center}
\end{figure}
The analytical result for the scaling function $\tilde f_\parallel
(\etatil)$, shown as the solid curve in Fig.\ref{fig:f_universal}, has
several characteristic features. First of all, it is always
monotonically increasing since the free energy is convex. For $\etatil
\gtrapprox 0.4$, the scaling function is $\tilde f_\parallel \approx
1$ corresponding to $f_\parallel \approx f_c$, i.e. a vanishing
contribution of thermal fluctuations to the force. For smaller
$\etatil$, corresponding to larger distances $\zeta$ between the wall
and the grafted end of the polymer, fluctuations reduce the force
exerted on the wall by effectively shortening the polymer. For $\zeta
\to L$ (resp.  $\etatil = 0$), the probability of the polymer to
contact the wall becomes smaller and smaller until finally for $\zeta
= L$ only one configuration, namely the completely straight one, has
$r_z(L) = L$.  Hence the force must vanish for all $\zeta \ge L$
(resp.  $\etatil \le 0$). 

We have learned already in Section \ref{sec:P_parallel} that there are
excellent approximations to the scaling function for the probability
density of the free polymer end for small values of the reduced stored
length, \eqref{eq:app_tip_distribution_3d_smaller}. In the same way,
the first term of \eqref{eqn:Z_parallel_smallrho} is an excellent
approximation to the infinite series for $\etatil \lessapprox 0.2$.
Thus we may write for the scaling function of the entropic force
\begin{eqnarray}
  \label{eqn:f_form3}
  \tilde f_\parallel^<(\etatil)
  = \frac{4 \, \e^{-1/4 \etatil}}{\pi^{5/2} {\etatil}^{3/2}
    \left(1 - 2 \, \erfc(1/2\sqrt{\etatil})\right)} \, , 
\end{eqnarray}
which already describes most of the nontrivial shape of the scaling
function. For $\etatil \gtrapprox 0.2$ it suffices to high accuracy
to use the first two terms of \eqref{eqn:Z_form1}, which gives
\begin{eqnarray}
  \label{eqn:f_form4}
  \tilde f_\parallel^>(\etatil)
  = \frac{1 - 3\e^{-2 \pi^2 \etatil}}
         {1 - \frac{1}{3}\e^{-2\pi^2 \etatil}}\,.
\end{eqnarray}
Upon inspection of \eqref{eq:entropic_force-best:3d} one may interpret
the functional form of the entropic force as due to two effects. In
the numerator we have the probability density for the position of the
free end at the wall. This function shows a pronounced peak as one
decreases the distance $\zeta$ (resp. increases the scaling variable
$\etatil$). At the same time, the denominator, the cummulative
distribution function, decreases with decreasing $\zeta$. It is now a
matter of how fast these changes occur what the ensuing shape of the
scaling function for the entropic force will be. In the present case
of a polymer in 3d the decrease in the cummulative distribution
function seems to be fast enough to compensate the maxiumum in the
probability density of the free polymer end such that the entropic
force becomes a monotonically increasing function of $\etatil$.

>From Fig.\ref{fig:f_universal} one observes that the universal scaling
curve is a lower bound to the MC data for all values of the stiffness
parameter $\varepsilon$. For fixed $\varepsilon$ the entropic force
always increases monotonically with increasing compression; for
intermediate values $\varepsilon \approx 2.5$ there is a pronounced
change in curvature at $\etatil \approx 0.25$. For strong compression
the results asymptote to the mechanical limit ($\kT = 0$). This limit
is not correctly reproduced within the harmonic approximation which
gives 
\begin{eqnarray}
  f_\text{mech} (\zeta)
  = f_c \Theta (L-\zeta) \, ,
\end{eqnarray}
whereas the exact force-extension curve is a monotonous function
in $\zeta$ that is somewhat larger than $f_c$ for $\zeta <
L$ and tends to $f_c$ for $\zeta\to L$. 

One might finally ask, whether these entropic forces $f_\parallel
(\zeta)$ are related to the force extension relation discussed in
section \ref{sec:mgf}, $\avg{r_z(L)}_f - \avg{r_z(L)}_0 =
k_\parallel^{-1} f + \order {(f^2)}$ with $k_\parallel = 6\kappa^2/\kT
L^4$ \cite{kroy-frey:96}. Rewriting these linear response result in
scaling form we find,
\begin{eqnarray}
  \frac{f}{f_c} = \frac{24}{\pi^2} \left(\etatil - \frac12 \right)\,.
\end{eqnarray}
Comparing this with Fig.\ref{fig:f_universal}, we see that the linear
response result does not contain any information about the situation
under investigation here. To the contrary, the initial rise of the
force when $\zeta$ becomes slightly smaller than $L$ is highly
nonlinear (see Eq. (\ref{eqn:f_form3})).

\subsection{Distribution function and entropic forces: 2d}
\label{sec:P_parallel2d} 

\begin{figure}[h]
  \begin{center}
    \psfrag{x}{\hspace{-1cm}$\rhotil=(L-z)/L_{\parallel}$}
    \psfrag{y}{\hspace{-0.3cm}$\tilde P_\parallel(\rhotil)$}
    \rotatebox{-90}{\includegraphics[height=\columnwidth]
      {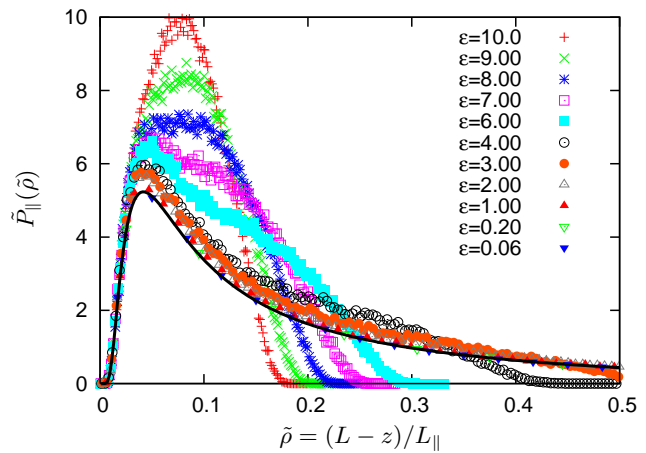}}
    \caption{Probability density $\tilde P_\parallel(\rhotil)$ of 
      the free end of a grafted semiflexible polymer in $2d$ as a
      function of $\rhotil$ (solid curve). MC data for different
      stiffness parameters as indicated in the graph.}
    \label{fig:P_parallel2d}
  \end{center}
\end{figure}

Analogous to the previous section the tip distribution function of a
polymer confined to 2d, e.g. by two parallel glass plates, obeys a
scaling law in the stiff limit
\begin{eqnarray}
P_\parallel (z,L,\lp) = \lpar^{-1} \tilde P_\parallel (\rhotil) \, .
\end{eqnarray}
The scaling function may again be represented in terms of series
expansions (see Appendix \ref{app:inverse_laplace2d}). A series which
converges well for small values of $\rhotil$ reads
\begin{eqnarray}
\label{eqn_pofz2d}
 \tilde P_\parallel (\rhotil) 
 =  \sum_{l=0}^{\infty} 
         \left(\begin{array}{c}  
           - \frac12 \\ l 
         \end{array} \right)
     \frac{2l+\frac12}{\sqrt{2\pi} \, \rhotil^{3/2}}
 \exp{\left[-\frac{(l+\frac14)^2}{\rhotil}\right]}
 \, ;
\end{eqnarray}
for an explicit formula for the binomial coefficient in
\eqref{eqn_pofz2d} see \eqref{eq:binomial}.  The scaling function,
shown as the solid curve in Fig.\ref{fig:P_parallel2d}, has an overall
shape which is quite similar to 3d with a pronounced maximum close to
full stretching.
The series approximations may again give useful approximate
expressions for the shape. In the proximity of full stretching, the
series given by \eqref{eqn_pofz2d} converges very fast such that
already the first term
\begin{eqnarray}
 \tilde P^<_\parallel (\rhotil) 
 = \frac{1}{\sqrt{8 \pi} \rhotil^{3/2}} 
   \exp \left( - \frac{1}{16 \rhotil} \right)
\label{eq:saddle_2d}
\end{eqnarray}
is an excellent approximation for the whole series at least for
$\rhotil \leq 0.3$. As in 3d, a saddle point approximation also gives
\eqref{eq:saddle_2d} (see Appendix \ref{app:saddle_point}).
Alternatively, as shown in Appendix \ref{app:inverse_laplace2d}, one
may derive a series expansion which converges well in the strong
compression limit; see \eqref{eqn_pofz2dSecondForm}. For $\rhotil
\gtrapprox 0.3$ it suffices to use the first term of this sum only
which reads
\begin{eqnarray}
 \tilde P^>_\parallel(\rhotil)
 &=& \frac{\pi \e^{-\pi^2\rhotil/4}}{2\sqrt{2}}
 \left[ 1+1.5\,e^{-5\pi^2\rhotil/16} +2\, \e^{-12\pi^2\rhotil/16} \right.
 \nonumber\\
 &&+
 \left. 2.5\,\e^{-21\pi^2\rhotil/16}+3\,\e^{-32\pi^2\rhotil/16}\right]
 \, .
\end{eqnarray}

Upon increasing the stiffness parameter the rescaled probability
distribution deviates from the scaling function in the semiflexible
limit and approaches a Gaussian. In contrast to 3d there is an
intermediate parameter regime in the stiffness parameter where $\tilde
P_\parallel (\rhotil)$ exhibits a marked shoulder. This feature of the
distribution function has recently been identified and explained in terms of an
interesting analogy with the physics of a random walker in shear flow
\cite{benetatos-munk-frey:2005}.

\begin{figure}[htbp]
\begin{center}
  \psfrag{x}{\hspace{-1cm}$\etatil=(L-\zeta)/L_{\parallel}$}
  \psfrag{f}{\hspace{-0.3cm}$\tilde F_\parallel(\etatil)$}
  \rotatebox{-90}{\includegraphics[height=\columnwidth]
    {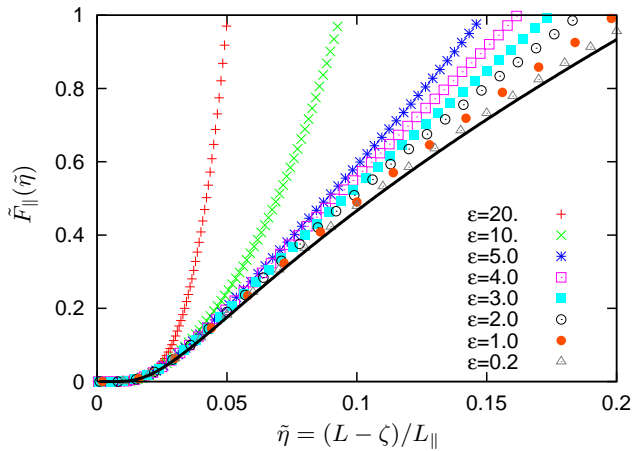}}
\caption{Free energy of a grafted polymer whose tip is  
  constrained by a rigid wall in 2d. The solid line gives the
  universal scaling function in the stiff limit.  MC data are
  given by the symbols for different values of the stiffness parameter
  $\varepsilon$ as indicated in the graph.}
\label{fig:FreeEnergy2d_Universal}
\end{center}
\end{figure}

\begin{figure}[h]
\begin{center}
  \psfrag{x}{\hspace{-1cm}$\etatil=(L-\zeta)/L_{\parallel}$}
  \psfrag{f}{\hspace{-0.5cm}$\tilde f_\parallel(\etatil)/f_c$}
  \rotatebox{-90}{\includegraphics[height=\columnwidth]
    {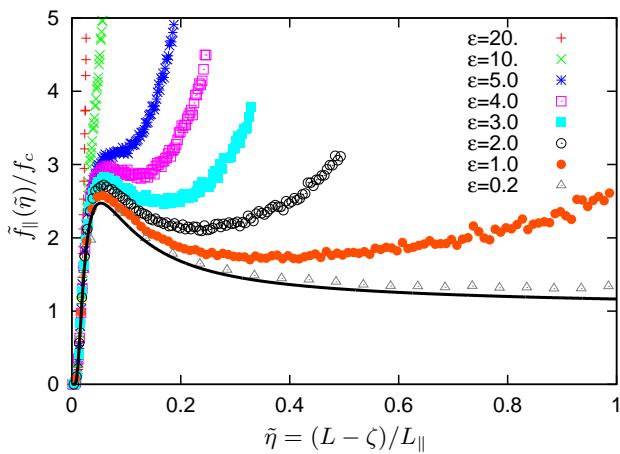}}
 \caption{Scaling function for the entropic force which a grafted polymer 
   exerts on a rigid wall in 2d as a function of the reduced stored
   length $\etatil = (L-\zeta)/\lpar$. The solid line gives the
   universal scaling function in the stiff limit.  MC data are
   given by the symbols for different values of the stiffness
   parameter $\varepsilon$ as indicated in the graph.}
\label{fig:Force2d_Universal}
\end{center}
\end{figure}

Upon integrating \eqref{eqn_pofz2d} from $-L$ to $\zeta$ one obtains
for the restricted partition sum
\begin{eqnarray}
\label{eqn:partition_2d} 
 {\cal Z}_\parallel (\zeta) 
 = 1 -\sqrt 2 \sum_{k=0}^{\infty} 
   \frac{(-1)^k (2k\!-\!1)!!}{2^k k!}
   \erfc \left( \frac{\lambda_{2k+1}}{2 \pi \sqrt{\etatil}}\right)
   \, , 
\end{eqnarray}
with the same scaling variable $\etatil$ as in the previous section.
Similarly, using \eqref{eqn_pofz2dSecondForm} gives
\begin{eqnarray}
\label{eqn:partition_2dSecondForm}
  {\cal Z}_\parallel (\zeta) \approx 
 \frac{1}{1.49} \sum_{k=0}^{\infty} (-1)^k
 \sum_{i=4}^8 \lambda_{2k+i/4}^{-1} 
 \e^{-\lambda_{2k+i/4}^2\etatil} \, .
\end{eqnarray}
Hence, as in 3d, one finds for the free energy
\begin{eqnarray}
 {\cal F}_\parallel (\zeta, L, \lp) 
  = - \kT \ln \tilde {\cal Z}_\parallel (\etatil)
\end{eqnarray}
and the entropic force 
\begin{eqnarray}
  f_\parallel (\zeta, L, \lp) = f_c \, \tilde f_\parallel (\etatil) 
\end{eqnarray}
with the scaling function 
 \begin{eqnarray}
 \tilde{f}_\parallel (\etatil) 
  = -\frac{4}{\pi^2} 
   \frac{\tilde{\cal Z}'_\parallel (\etatil)}
        {\tilde{\cal Z}_\parallel (\etatil)}
\end{eqnarray}
and $f_c= \pi^2 \kappa / 4 L^2$; see the solid curves in
Fig.\ref{fig:FreeEnergy2d_Universal} and
Fig.\ref{fig:Force2d_Universal} for a plot of the scaling functions
for the free energy and entropic force, respectively.
The key difference between the results in 2d and 3d is that the
effective free energy exhibits a change in curvature at $\etatil
\approx 0.05$ and as a result a pronounced peak in the entropic force.
The peak is a pretty robust feature of the distribution function and
vanishes only for very large values of $\varepsilon \approx 5$.

In order to understand the physical origin of this peak it suffices to
consider small values of $\etatil$. Then, using only the leading term
of the series expansion \eqref{eqn:partition_2d}, one obtains for the
entropic force
\begin{eqnarray}
\label{eqn:forceform2_2d} 
  \tilde f_\parallel^<(\etatil)
  = \frac{\sqrt2 \e^{-1/16\etatil}}
    {\pi^{5/2}\etatil^{3/2}\left(1-\sqrt 2
    \erfc[1/4\sqrt{\etatil}]\right)} \, .
\end{eqnarray}
This has the same functional form as the corresponding expression in
3d, \eqref{eqn:f_form3}, but differs in some numerical factors.  These
differences can all be traced back to the strength $\alpha$ of the
essential singularity of the tip distribution function close to full
stretching, $\tilde P_\parallel (\rhotil) \propto \exp (-\alpha /
\rhotil)$; compare \eqref{eq:app_tip_distribution_3d_smaller} with
\eqref{eq:saddle_2d}.  One may interpret this strength as a kind of
phase space factor counting how fast the number of polymer
configurations decreases as one approaches full stretching. It clearly
shows that the maximum of the entropic force in 2d is of purely
geometric origin. As an interesting consequence of this maximum one
should note, that for most values of the reduced stored length
$\etatil$ the entropic force {\em exceeds} the purely mechanical
force given by the Euler buckling force.

\section{Grafted polymer at an oblique angle to the wall}
\label{sec:oblique_angle}

\begin{figure}[htbp]
  \begin{center} 
    \psfrag{x}{$x$} \psfrag{z}{$z$} \psfrag{t}{$\vartheta$}
    \psfrag{u}{$\zeta$} \psfrag{n}{$\hat n$} \includegraphics[width=6
    cm]{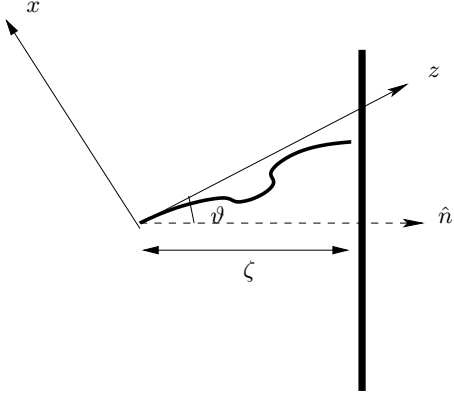}
        \caption{A smooth hard wall at some oblique angle $\pi/2 - \vartheta$
        constrains the configurations of a stiff polymer grafted parallel
        to the $z$-axis.}
         \label{fig:2d_cartoon}
  \end{center}  
\end{figure}

The generic situation one encounters in a cellular system is that the
polymer is inclined with respect to a membrane. Then we have to ask
how the force derived above changes when the graft of the polymer is
not orthogonal to the constraining wall but at some oblique angle
$\pi/2 - \vartheta$; see Fig.\ref{fig:2d_cartoon}. Since the presence of
the wall restricts the position of the polymer tip to
\begin{eqnarray}
  r_z (L) \cos \vartheta + r_x (L) \sin \vartheta \leq \zeta \, 
\end{eqnarray}
one has to evaluate the restricted partition sum
\begin{eqnarray}
  \label{eqn:Z_general_def}
  {\cal Z} (\zeta, \vartheta)
    &=& \avg{\Theta[\zeta \!-\! r_z(L) 
             \cos \vartheta \!-\! r_x(L) \sin \vartheta]}_0
    \nonumber \\
    &=& \int dx dz P(x,z) \,
        \Theta[\zeta \!-\! z \cos\vartheta \!-\! x \sin\vartheta] \, 
\end{eqnarray}
to find the entropic force. 

\subsection{Probability distribution function of the tip}

This calculation requires the knowledge of the joint probability
density of the tip
\begin{eqnarray}
  P(x, z) := \avg{\delta[r_x(L) - x]
                  \, \delta[r_z(L) - z]}_0 \,.
\end{eqnarray}

In Section \ref{sec:orthogonal} we have already analyzed the reduced
distribution function $P_\parallel (z)$ and found that its width is
characterized by the scale $\lpar=L^2/\lp$. Similarly, one can find an
explicit expression for $P_\perp (x)$ in harmonic approximation, where
\begin{eqnarray}
  \label{eqn:rx_by_a}
  r_x(L) &\approx& \sum_{k=1}^\infty a_{x, k} \int_0^L d  s
  \, \sin (\lambda_k s / L)
  \nonumber \\
  &=& L \sum_{k=1}^\infty \lambda_k^{-1} a_{x, k} \, ,
\end{eqnarray}
and thus
\begin{eqnarray}
  P_\perp(x)
  &=& \int \frac{d  q}{2\pi} \, \e^{\i q x}
  \left\langle \exp \left[ 
  -\i q L \sum_{k=1}^\infty \lambda_k^{-1} a_{x, k}
  \right] \right\rangle
  \nonumber \\
  &=& \int \frac{d  q}{2\pi} \, \e^{\i q x}
  \exp \left[- \frac{L^3}{\lp} q^2\sum_{k=1}^\infty \lambda_k^{-4}
       \right] \,.
\end{eqnarray}
With $\sum_{k=1}^\infty \frac{1}{(2k-1)^4} = \frac{\pi^4}{96}$
\cite{Stegun}, this gives a Gaussian distribution
\begin{eqnarray}
  P_\perp(x) = \frac{1}{\sqrt{2\pi}\lperp}
               \e^{-\frac12 (x/\lperp)^2} \, ,
\label{eq:transverse_tip_distribution}
\end{eqnarray}
where we have defined the characteristic {\em transverse length scale} 
\begin{eqnarray}
\lperp = \sqrt{L^3/3 \lp} \, .
\end{eqnarray}
Together with $\lpar$ these are the two length scales are
characterizing the width of the joint distribution function. This
suggests to write the joint distribution function as
\begin{eqnarray}
  P (x,z,L,\lp) = \frac{1}{\lpar \lperp} 
            \tilde P (\tilde x, \rhotil)
\end{eqnarray}
in terms of dimensionless variables 
\begin{eqnarray}
\tilde x &=& x / \lperp \, , \\
\rhotil &=& (L-z)/\lpar \, .
\end{eqnarray}

An explicit form of the joint distribution function can be calculated
to harmonic order. For simplicity, we start with a polymer fluctuating
only in the $x$-$z$-plane ($d=2$).  Then
\begin{widetext}
\begin{eqnarray}
  \label{eqn:P_calc_1}
  P_2(x, z) &=& \int \frac{d q_z}{2\pi} \frac{d q_x}{2\pi} \e^{-\i q_z z
    - \i q_x x} \avg{\e^{\i (q_z r_z(L) + q_x r_x(L)) }}_0
  \nonumber \\
  &=& \int \frac{d q_z}{2\pi} \frac{d q_x}{2\pi} \, 
           \e^{\i q_z (L-z) - \i q_x x } \, 
           \prod_k 
           \left\langle 
           \exp\left[-\i \left(\frac{L q_z}{4} a_{x,k}^2 - 
                               \frac{L q_x}{\lambda_k} a_{x,k}\right)
               \right]
           \right\rangle
  \nonumber \\
  &=& \int \frac{d q_z}{2\pi}\frac{d q_x}{2\pi} \e^{\i q_z (L-z) - \i
    q_x x } \prod_k \sqrt{\frac{\lambda_k^2} {\lambda_k^2 + \i q_z
      \lpar}} \exp\left[-\frac{3 q_x^2 \lperp^2}{\lambda_k^2
      (\lambda_k^2 + \i q_z \lpar)}\right]
  \nonumber \\
  &=& \frac{1}{\lperp \lpar} \int \frac{d \tilde{q}_z}{2\pi}\frac{d
    \tilde{q}_x}{2\pi} 
    \e^{\i \tilde{q}_z \rhotil - \i \tilde{q}_x \tilde x} \left(\prod_k
    \sqrt{\frac{1}{1 + \i \tilde{q}_z \lambda_k^{-2}}}\right) \exp\left[- 3
    \tilde{q}_x^2 \sum_k 
    \frac{1}{\lambda_k^2 (\lambda_k^2 + \i \tilde{q}_z)}\right]
  \nonumber \\
  &=& \frac{1}{\lperp \lpar} \int \frac{d \tilde{q}_z}{2\pi}\frac{d
    \tilde{q}_x}{2\pi} a_2(\i \tilde{q}_z) 
    \e^{\i \tilde{q}_z \rhotil - \i \tilde{q}_x \tilde x}
    \exp\left[-\frac{3}{2}~\tilde{q}_x^2~b(\i \tilde{q}_z)\right]\,,
\end{eqnarray}
\end{widetext}
where for $z \in \mathbb{R}_+$ we have \cite{Stegun}
\begin{eqnarray}
  a_2(z)&:=& \prod_k \sqrt{\frac{1}{1 + z \lambda_k^{-2}}}
         = \sqrt{\frac{1}{\cosh \sqrt{z}}} \, ,\\
  b(z)  &:=&2 \sum_k\frac{1}{\lambda_k^2 (\lambda_k^2  + z)} 
         = \frac{\sqrt{z}-\tanh\sqrt z}{z^{3/2}} \,.
\end{eqnarray}
For $d=3$, the additional degrees of freedom associated with
excursions in the $y$-direction lead to the replacement of $q_z
a_{x}^2(k)$ by $q_z [a_{x}^2(k) + a_{y}^2(k)]$ which results in an
additional factor of $\sqrt{1 + \i q_z \lambda_k^{-2}}$ for each mode
$k$. Thus, for general $d$, we have to replace $a_2(z)$ with
\begin{eqnarray}
  a_d(z) := 
 \prod_k\left[ \frac{1}{1 + z \lambda_k^{-2}} \right]^{(d-1)/2}.
\end{eqnarray}

As $\Re \left[ b(\i \tilde{q}_z) \right] > 0$ for all
$\tilde{q}_z \in [-\infty,\infty]$, the Gaussian integration over
$\tilde{q}_x$ in  Eq.(\ref{eqn:P_calc_1}) can be performed by completing the square, such that
\begin{eqnarray}
  \label{eqn:Pfull}
  \tilde P_d (\tilde x, \rhotil)
  = \int \frac{d  \tilde{q}_z}{2\pi}
      \e^{\i \tilde{q}_z \rhotil}
      \frac{a_d(\i \tilde{q}_z)}{\sqrt{6\pi  b(\i \tilde{q}_z)}}
      \exp\left[-\frac{\tilde x^2}{6 b(\i \tilde{q}_z)}\right] \, .
\end{eqnarray}

Along similar lines one may also calculate the full joint distribution
function for a grafted polymer in $d=3$,
\begin{eqnarray}
\label{eqn:PfullXYZ}
  P_3 (x, y, z)
  &=& \frac{1}{\lperp^2\lpar} \int \frac{d  \tilde{q}_z}{2\pi}
      \e^{\i \tilde{q}_z \rhotil}
      \frac{a_3(\i \tilde{q}_z)}{6\pi  b(\i \tilde{q}_z)}
      \exp\left[-\frac{\tilde x^2+\tilde y^2}{6 b(\i \tilde{q}_z)}\right]
  \nonumber \\
  &=:& \frac{1}{\lperp^2
    \lpar} \tilde P_3 (\tilde x, \tilde y,\rhotil) \,.
\end{eqnarray}
In addition to the poles of $a_3(\i \tilde{q}_z)$ at $\tilde{q}_z = \i
\lambda_k^2$ on the positive imaginary axis of the
$\tilde{q}_z$-plane, the integrand also has singularities at the zeros
$\i \lambda_k^2$ of $b(z)$. Thus we continue by evaluating the
integrals numerically.

\subsubsection{Numerical evaluation of integrals}

The integrand of \eqref{eqn:Pfull} has no singularities on the real
$\tilde q_z$-axis.  Before attempting a numerical integration, we discuss the
behavior of the different terms appearing in \eqref{eqn:Pfull}. For
$d=3$, we have
\begin{eqnarray}
  a_3(z) = \prod_k \frac{1}{1 + z \lambda_k^{-2}} =
  \frac{1}{\cosh \sqrt{z}} \,.
\end{eqnarray} 
For $\tilde q_z \in \mathbb R$, the real and the imaginary part
$1/\cosh \sqrt{\i \tilde q_z}$ are respectively even and odd functions
rapidly decaying in magnitude for $\tilde q_z \to \pm\infty$. The real
part of $1/b(\i \tilde q_z)$ is strictly positive and increasing with
increasing $|\tilde q_z|$. The imaginary part of $1/b(\i \tilde q_z)$
behaves asymptotically as $\Im \left[ b^{-1}(\i \tilde q_z) \right]
\sim \tilde q_z$ leading to a second strongly oscillating contribution
to the integrand of \eqref{eqn:Pfull} besides $\exp (\i q z)$. In the
interest of numerical stability of the integration, it is advantageous
to rewrite the integrand appearing in \eqref{eqn:Pfull} to
\begin{eqnarray}
  \label{eqn:Pfull_num}
  \frac{1}{2\pi}
  \e^{\i q (\rhotil - \tilde x^2/6)}\frac{a_3(\i q)}
  {\sqrt{2 \pi 3 b(\i q)}}
  \exp\left[- \frac{\tilde x^2}{6}\left(1/b(\i q) - \i q\right)\right]
\end{eqnarray}
for $q$ larger than some fixed $q_0$.

\subsubsection{Region of vanishing probability}
\label{sec:vanishing_prob}

\eqref{eqn:Pfull_num} suggests that $\rhotil = \tilde x^2/6$ is a
special situation.  The probability density $P(x,z)$ must vanish for
points which are at distances greater than $L$ from the graft: $x^2+z^2
\le L^2$. What does this translate to in the harmonic approximation?
The largest value $x^*$ of $r_x(L)$ that can be obtained for a given
value $z^*$ of $r_z(L)$ can be found from the variation of $r_x(L) -
\rho (z^* - r_z(L))$ where $\rho$ is a Lagrange multiplier.  Using
\eqref{eqn:rz_by_a} and \eqref{eqn:rx_by_a}, this leads to $a_{x, k} =
a/\lambda_k$ where $a$ is some number.  We thus find
\begin{eqnarray}
  x^* = L a \sum_{k=1}^\infty \lambda_k^{-2} = L \frac{a}{2}
\end{eqnarray}
and
\begin{eqnarray}
  \label{eqn:vanish_cond}
  z^* = L - \frac{L}{4} a^2 \sum_{k=1}^\infty
  \lambda_k^{-2} = L - L \frac{a^2}{8}
\end{eqnarray}
resulting in
\begin{eqnarray}
  \frac{L - z^*}{L} = \frac{1}{2}
  \left(\frac{x^*}{L}\right)^2 \, .
\end{eqnarray}
As $\lperp^2/L \lpar = 1/3$, this is equivalent to
\begin{eqnarray}
  \rhotil^* = \frac{1}{6} (\tilde x^*)^2 \,.
\end{eqnarray}
Hence $\tilde P(\tilde x, \rhotil)$ has to vanish for $\rhotil <
\tilde x^2/6$ .

\subsubsection{Results for the general distribution function}

\begin{figure}[hbtp]
  \begin{center}
    \psfrag{x}{$\tilde x$} \psfrag{z}{$\rhotil$}
    \includegraphics[width=7.5 cm]{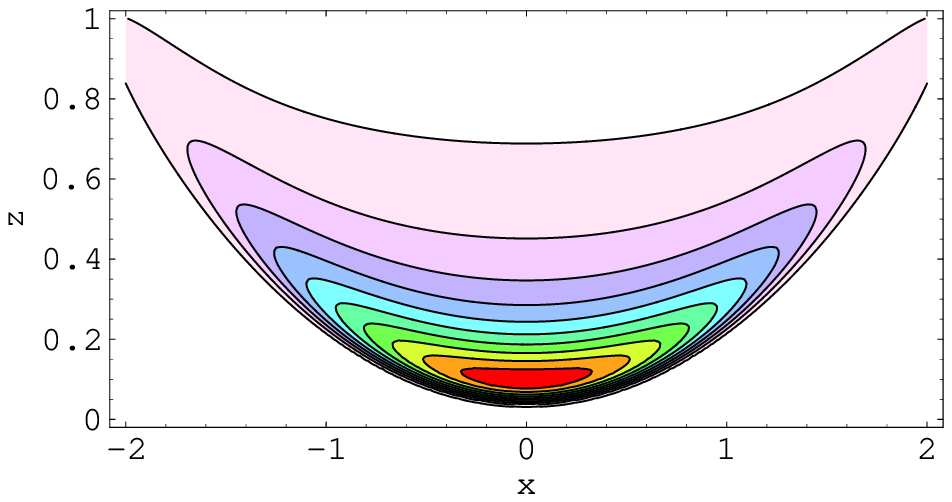} 
    \includegraphics[width=7.5 cm]{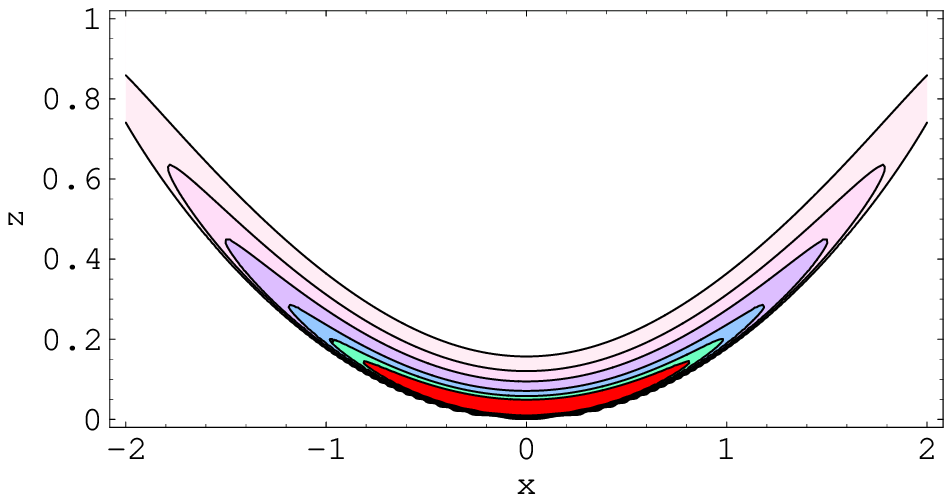}
    \caption{Contour plot of the probability density $\tilde
      P(\tilde x, \rhotil)$ in $d=3$ (top) and $d=2$ (bottom).}
    \label{fig:ContourPlot}
  \end{center}
\end{figure}

It is now straightforward to evaluate the integrals in
Eq. (\ref{eqn:Pfull}) by some standard numerical method. The
corresponding results are shown in Fig.\ref{fig:ContourPlot} as
contour plots of $\tilde P(\tilde x, \rhotil)$ in $d=3$ and $d=2$,
respectively. 
\begin{figure}[hbtp]
  \begin{center}
    \psfrag{x}{$\tilde x$} \psfrag{y}{$\rhotil$}
    \rotatebox{-90}{\includegraphics[height=8cm]
    {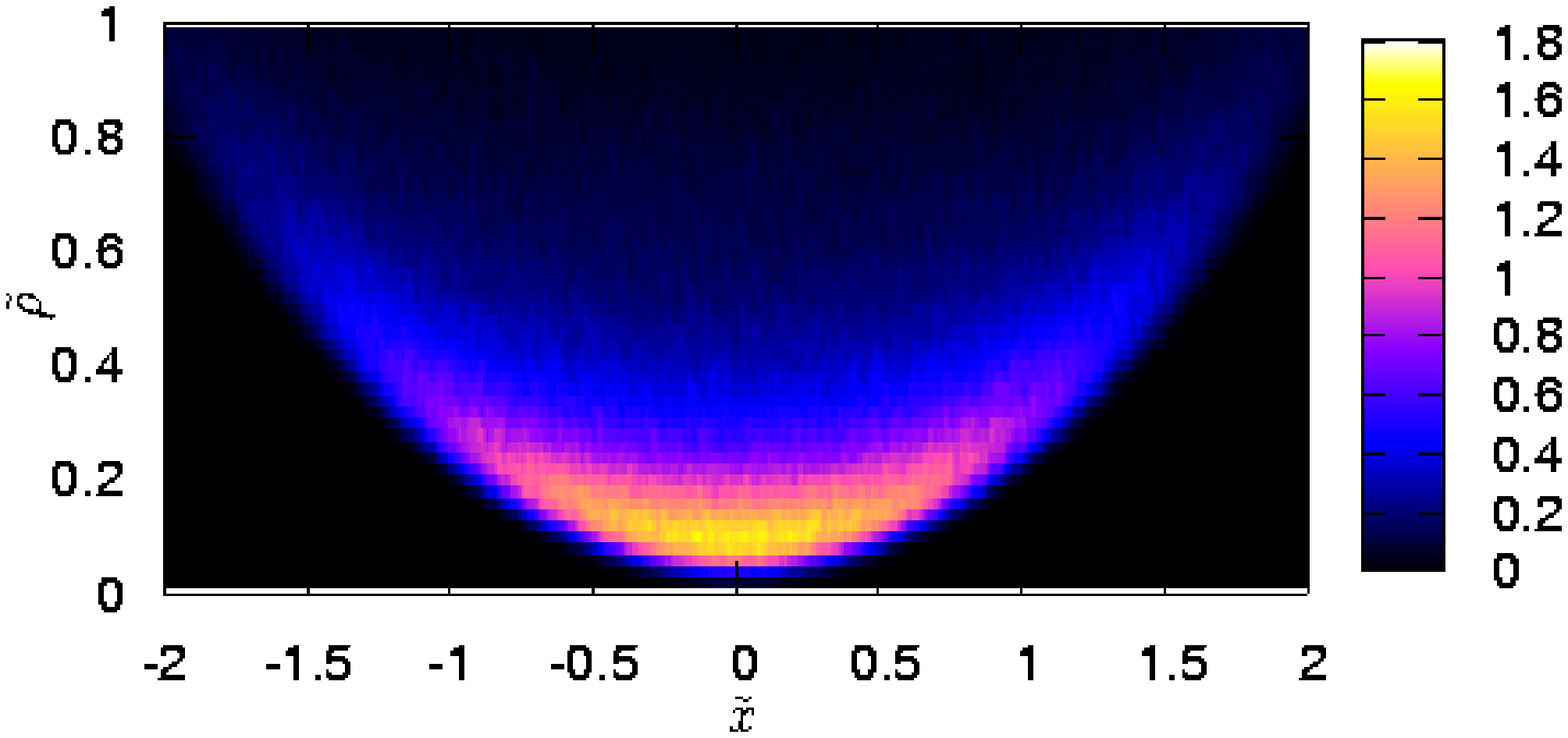}}
    \rotatebox{-90}{\includegraphics[height=8cm]
    {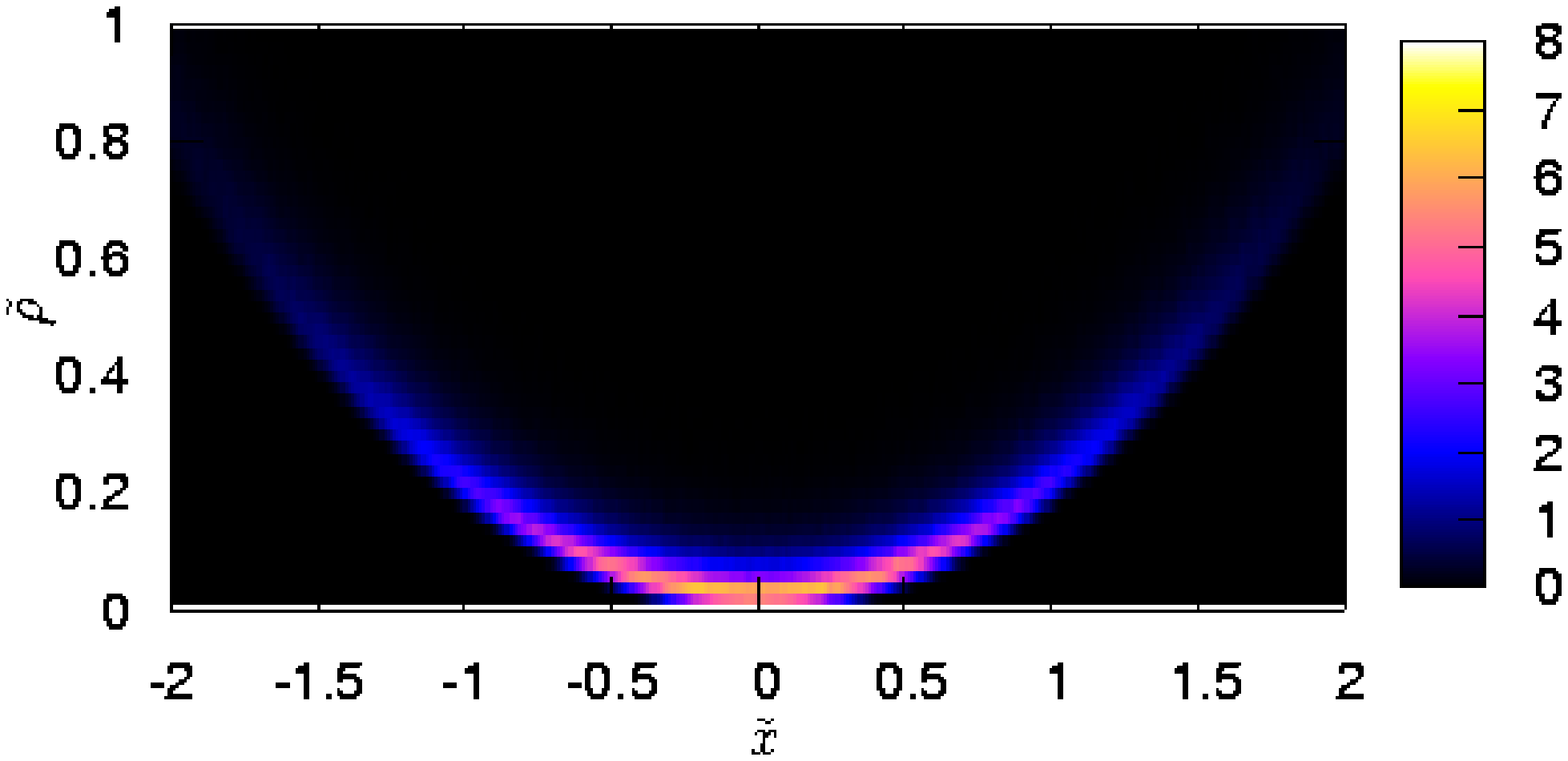}}
    \caption{Color coded density plot of the probability density 
      $\tilde P(\tilde x, \rhotil)$ in $d=3$ (top) and $d=2$ (bottom)
      obtained from MC simulations for $\varepsilon=0.1$.}
    \label{fig:ScaledDensityPlot}
  \end{center}
\end{figure}
These analytical results compare very well with MC results for
polymers with a stiffness parameter $\varepsilon \leq 0.2$; see a plot
with $\varepsilon = 0.1$ in Fig.\ref{fig:ScaledDensityPlot}.  There
are deviations between the harmonic approximation and MC data for
larger values of $\varepsilon$ \cite{lattanzi-munk-frey:2004,
  benetatos-munk-frey:2005, wagner-gholami-lattanzi-frey:06}.

The density distribution essentially vanishes outside the parabola
given by $\rhotil = \tilde x^2/6$, corresponding to the classical
contour of the polymer in harmonic order. The main weight of $\tilde
P(\tilde x, \rhotil)$ is concentrated close to this line, where the
effect is stronger for $d=2$. Profiles parallel to the $\rhotil$
direction are of a shape qualitatively similar to $\tilde
P_\parallel(\rhotil )$ (see Fig.\ref{fig:P_parallel}) at least for
small $\tilde x$.  Profiles parallel to the $\tilde x$-axis are not
Gaussian.  For small $\rhotil \lessapprox 0.1$, they are peaked at
$\tilde x = 0$ but unlike a Gaussian, they vanish for $\tilde x^2 > 6
\rhotil$. For larger $\rhotil$, they display a double-peaked shape.
Both features would be completely missed by a factorization
approximation $\tilde P(\tilde x, \rhotil) = \tilde P_\perp(\tilde
x)\tilde P_\parallel(\rhotil)$. An elaborate discussion of the
features of the distribution function in $d=2$ and $d=3$ as one
increases the stiffness parameters or introduces some backbone
elasticity will be the topic of a forthcoming publication
\cite{wagner-gholami-lattanzi-frey:06}.

\begin{figure}[h]
  \begin{center}
    \psfrag{x}{$\tilde x$} 
    \psfrag{y}{$\tilde y$} 
        \psfrag{z}{$\rhotil$}
     \includegraphics[width=0.8\columnwidth]{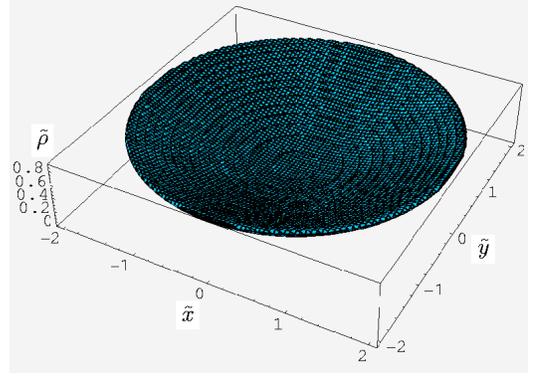}
    \caption{$3d$ probability isosurface $\tilde P_3(\tilde x, 
      \tilde y, \rhotil) = 0.1$ of a grafted polymer calculated
      numerically from Eq. (\ref{eqn:PfullXYZ}).}
    \label{fig:Isosurface3d}
  \end{center}
\end{figure}

The shape of the full joint probability distribution $P_3(x,y,z)$ is
best illustrated by plotting an isosurface, e.g. $\tilde P_3(\tilde x,
\tilde y, \rhotil)=0.1$ as shown in Fig.\ref{fig:Isosurface3d}. Due to
rotational symmetry a density plot for $P_3(x,y,z)$ may be obtained by
rotating the contour plot of $P_3(x,z)$ (Fig.\ref{fig:ContourPlot})
around the $z$-axis. Again MC and analytical results are identical for
small $\varepsilon$.

\subsection{Entropic forces: scaling functions}

\begin{figure}[htbp]
  \begin{center}
    \psfrag{1}{$\tan \alpha =\mu$} \psfrag{2}{$\rhotil = \tilde
      x^2/6$} \psfrag{Z}{$\eta_\parallel$} \psfrag{X}{$\eta_\perp$}
    \psfrag{x}{$\tilde x$} \psfrag{z}{$\rhotil$} \psfrag{a}{$\alpha$}
    \includegraphics[width=0.8\columnwidth]{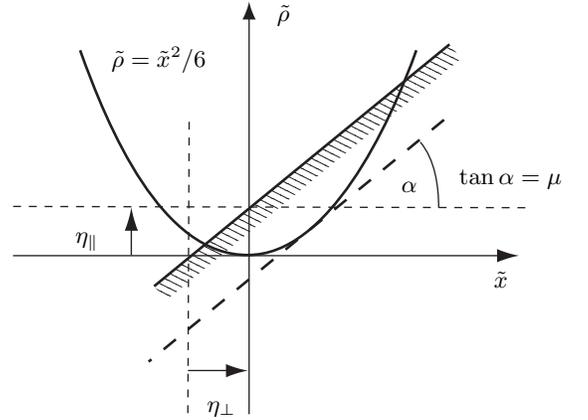}
   \caption{Geometry of the problem in terms of the reduced
    coordinates $\tilde x$ and $\rhotil$. The position of the wall is
    characterized by its slope $\mu=\eta_\parallel/\eta_\perp$ and
    $\eta_\parallel$, the distance from the origin along the $\tilde
    z$-axis, i.e. the minimal reduced stored length imposed by the
    presence of the constraining wall. To harmonic order the finite
    length of the filament also constrains the reduced stored length
    $\rhotil$ to be larger than $\tilde x^2/6$.}
    \label{fig:geometry_dimensionless_coordinate}
  \end{center}
\end{figure}

We are now in a position to evaluate the general expression
\eqref{eqn:Z_general_def} for the restricted partition sum.  Before
going into the details of the calculation it is instructive to have a
look at the geometry of the problem in terms of the dimensionless
variables $\tilde x$ and $\rhotil$. Recall that $\tilde x$ and
$\rhotil$ are measuring the transverse displacement of the tip $x$
and the stored length $L-z$ in units of the characteristic transverse
and longitudinal length scales, $\lperp$ and $\lpar$,
respectively.  As can be inferred from Fig.\ref{fig:2d_cartoon} the
wall crosses the $\tilde x$- and $\rhotil$-axis at
\begin{eqnarray}
 \eta_\perp     = \frac{L\cos\vartheta-\zeta}{\lperp\sin\vartheta} 
 \quad \text{and} \quad 
 \eta_\parallel = \frac{L\cos\vartheta-\zeta}{\lpar\cos\vartheta}
 \, , 
\end{eqnarray}
respectively; see Fig.\ref{fig:geometry_dimensionless_coordinate}.
These are the two basic dimensionless variables characterizing the
entropic forces exerted on the inclined wall. We also introduce the
slope $\mu = \tan \alpha$ of the constraining wall with respect to the
$\tilde x$-axis
\begin{eqnarray}
 \mu = \frac{\eta_\parallel}{\eta_\perp}
  = \frac{\lperp}{\lpar} \tan \vartheta 
  = \frac{1}{\sqrt{3 \varepsilon}} \tan \vartheta \,.
\label{eq:definition_mu}
\end{eqnarray}
As discussed above the finite length of the polymer gives a constraint
on the reduced stored length $\rhotil$ such that it has to be larger
than $\tilde x^2/6$, i.e. above the parabola drawn in
Fig.\ref{fig:geometry_dimensionless_coordinate}. Hence, just the
points on the constraining wall inside the parabola are accessible to
the tip of the polymer. As one moves the wall further away from the
grafted end, the number of contact points decreases and finally
reduces to zero when the wall becomes tangent to the parabola. In this
limit, where
\begin{eqnarray}
\eta_\parallel^c = - \frac{3}{2}\mu^2
\end{eqnarray}
the force exerted on the wall vanishes.

We may now write the restricted partition sum in terms of the reduced
stored length $\eta_\parallel$ and the slope of the wall $\mu$
\begin{eqnarray}
  {\cal Z}(\zeta, \vartheta) = \tilde{\cal Z} (\eta_\parallel, \mu) \, ,
\end{eqnarray}
where
\begin{eqnarray}
  \label{eqn:Z_force_num}
  &&\tilde{\cal Z}(\eta_\parallel, \mu) = \frac{1}{2} \erfc
  \frac{\eta_\parallel}{\sqrt{2}\mu} \\
  &&- \int_0^\infty
  \frac{d  q}{\pi q} \, \Im\left[\e^{\i q \eta_\parallel} \left(a_3(\i q)
    \e^{-\frac{3}{2} (\mu q)^2 b(\i q)}
  - \e^{-\frac12 (\mu q)^2}\right)\right] \, ,\nonumber
\end{eqnarray}
as shown in Appendix \ref{app:TechnicalDetailsofIntegrals}.  The force
is again found by taking the derivative of $\kT \ln {\cal Z}$ with
respect to $\zeta$. It obeys the scaling law
\begin{eqnarray}
  f(\zeta, \vartheta, L, \lp) = f_c (\vartheta) \,
  \tilde f \left(\eta_\parallel, \mu \right) \, , 
\end{eqnarray}
with an amplitude
\begin{eqnarray}
   f_c (\vartheta) =
  \frac{\pi^2 \kappa}{4 L^2\cos\vartheta} = \frac{f_c}{\cos\vartheta}
\end{eqnarray}
and a scaling function
\begin{eqnarray}
  \tilde f(\eta_\parallel, \mu)
  = -\frac{4}{\pi^2} \frac{\tilde{\cal Z}'(\eta_\parallel, \mu)}
                           {\tilde{\cal Z}(\eta_\parallel,\mu)} 
\label{eq:scaling_function_entropic_forces_general}
\end{eqnarray}
that can be expressed in terms of the restricted partition sum and its
derivative
\begin{eqnarray}
  \label{eqn:deriv_Z_force_num}
  \tilde {\cal Z}'(\eta_\parallel, \mu) = - \int_0^\infty
  \frac{d  q}{\pi} \Re\left(\e^{\i q \eta_\parallel} a_3(\i q)
    \e^{-\frac{3}{2} (\mu q)^2  b(\i q)}\right) \, .
\end{eqnarray}
As detailed in Appendix \ref{app:TechnicalDetailsofIntegrals}, Eq.
(\ref{eqn:Z_force_num}) and Eq. (\ref{eqn:deriv_Z_force_num}) are
suited best for a numerical evaluation of the entropic force. 

\begin{figure}[htbp]
  \begin{center}
    \psfrag{x}{\hspace{-1cm}$\eta_\parallel + 3\mu^2/2$}
    \psfrag{f}{$\tilde f$}
    \rotatebox{-90}{\includegraphics[height=\columnwidth]
      {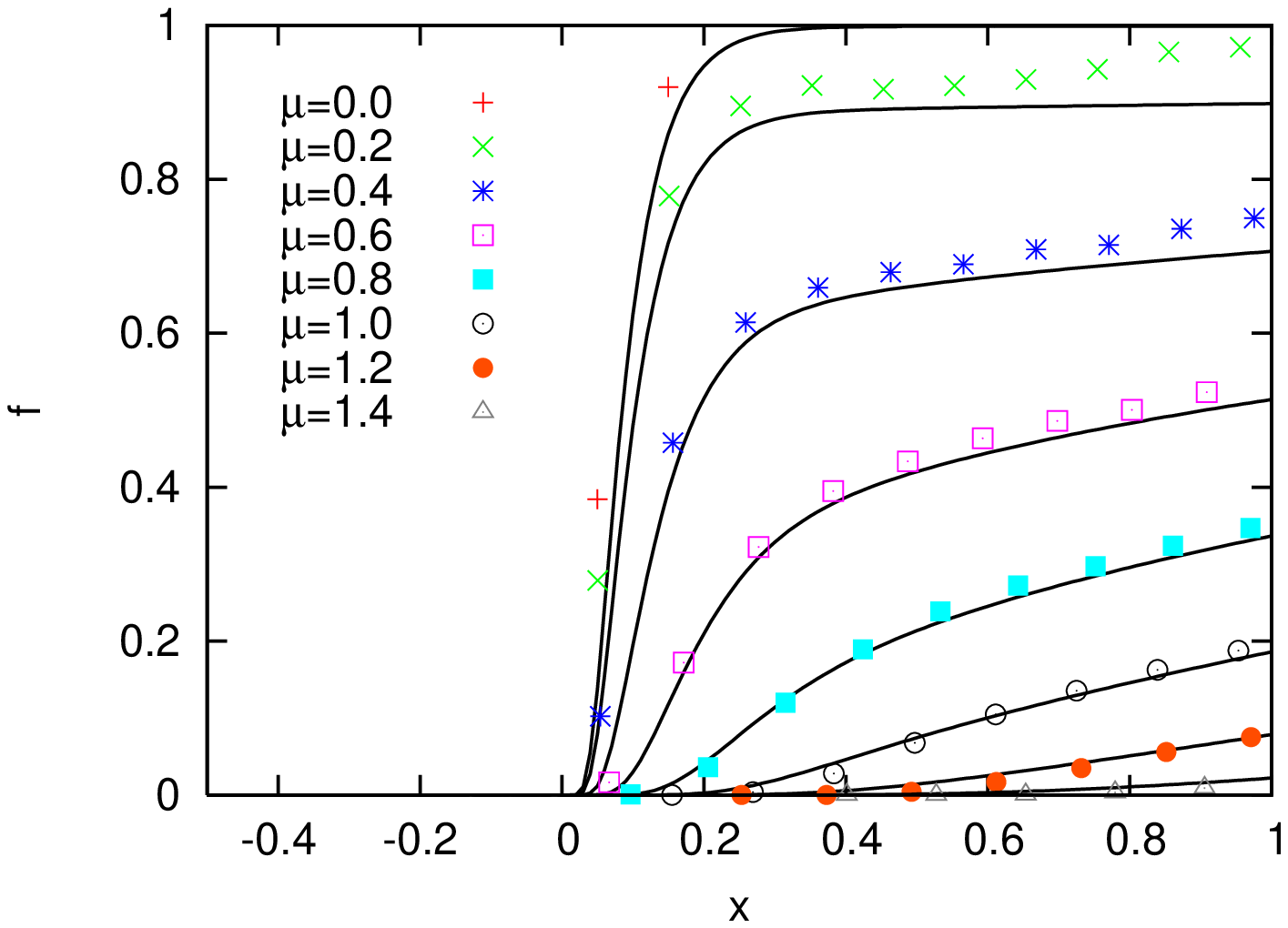}}
    \rotatebox{-90}{\includegraphics[height=\columnwidth]
      {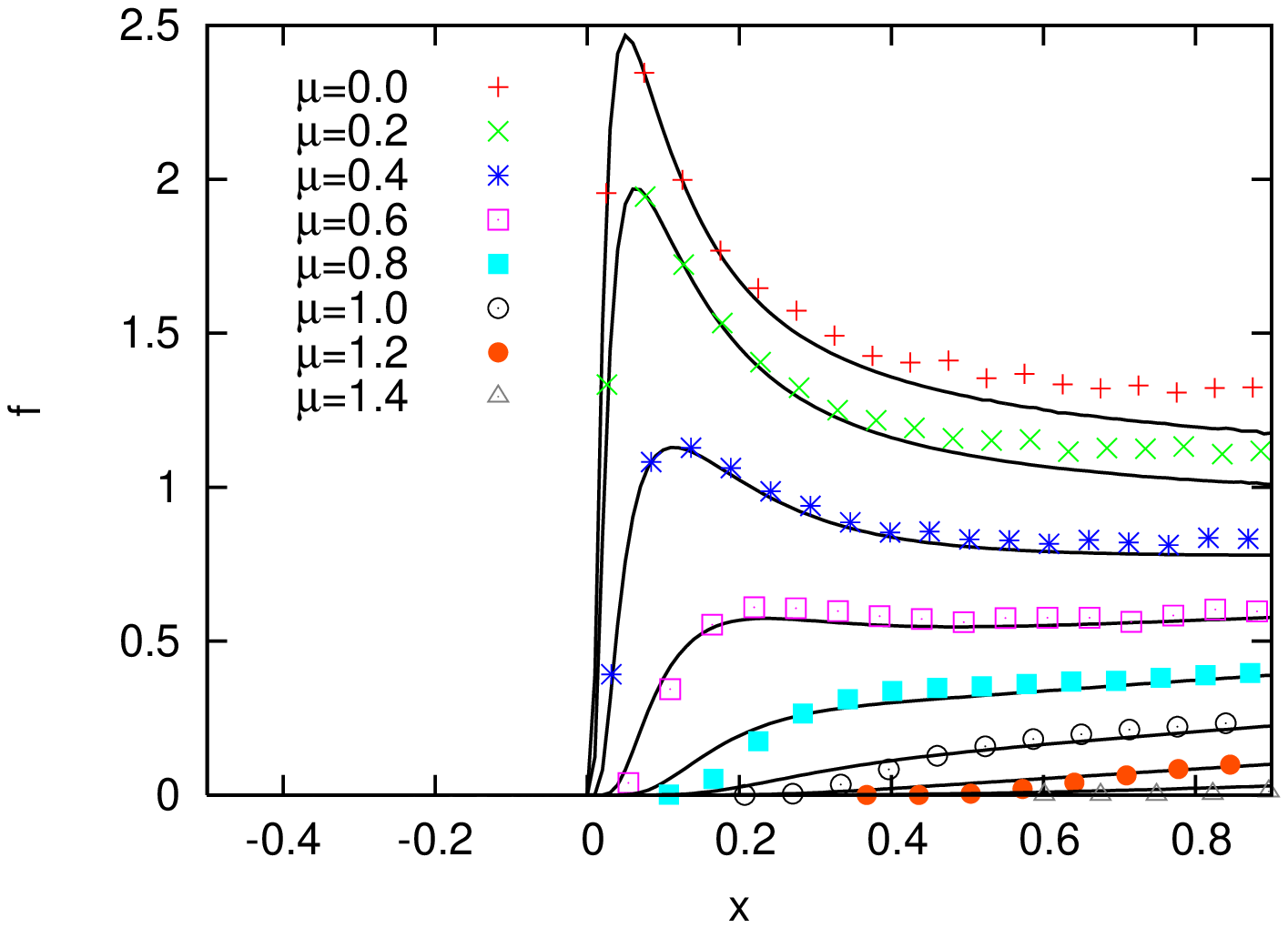}}
    \caption{Scaling function of the entropic force 
      $\tilde f$ in $d=3$ (top) and $d=2$ (bottom) as a function of
      $\delta \eta_\parallel = \eta_\parallel + \frac32 \mu^2$ for a
      series of values of $\mu$ as indicated in the graphs. Solid lines
      represent analytical results as obtained from a numerical
      evaluation of
      Eq.(\ref{eq:scaling_function_entropic_forces_general}). Monte
      Carlo data for a stiffness parameter $\varepsilon = 0.1$ are given
      as symbols as indicated in the graphs. For $\mu = 0$ one recovers
      the results for $\tilde f_\parallel(\eta_\parallel)$ as
      discussed in Section \ref{sec:orthogonal}.}
    \label{fig:fplot_parallel3dand2d}
  \end{center}
\end{figure}

In Fig.\ref{fig:fplot_parallel3dand2d} the analytical results for the
scaling function $\tilde f(\eta_\parallel, \mu)$ of the entropic
force are shown as a function of $\delta \eta_\parallel =
\eta_\parallel - \eta_\parallel^c$, for a series of values for
$\mu$.
Since we have subtracted off the critical value of the reduced stored
length $\eta_\parallel^c$, the forces vanish for $\delta
\eta_\parallel \leq 0$. There is a dramatic difference in the shape
of the force-distance curves in 2d and 3d. Whereas the force increases
monotonically with increasing $\delta \eta_\parallel$ for 3d, it
shows a pronounced maximum in 2d, the physical origin of which is the
same as for $\vartheta =0$. The maximum in 2d vanishes upon increasing
$\mu$, which can either be understood as an increase in the
inclination angle or an increase in the persistence length; see
Eq.(\ref{eq:definition_mu}).

For comparison MC data are given for a particular value of
the stiffness parameter, $\varepsilon = 0.1$. In this stiff regime the
analytical results compare very well with the MC data, except for
large values in the stored length where the harmonic approximation is
expected to become invalid.

For small values of $\mu$ the reduced stored length $\eta_\parallel$
is no longer a good variable. Instead we define a new scaling function
$\bar f (\eta_\perp, \mu)$ such that 
\begin{eqnarray}
  f(\zeta, \vartheta) = \frac{\kT}{\lperp \sin\vartheta} 
                     \bar f(\eta_\perp, \mu)
\end{eqnarray}
where
\begin{eqnarray}
  \bar f(\eta_\perp, \mu) = \mu \frac{\pi^2}{4}
  \tilde f(\eta_\parallel/\mu, \mu) \,.
\end{eqnarray}
\begin{figure}[htbp]
  \begin{center}
    \psfrag{x}{$\eta_\perp$} \psfrag{y}{${\bar f}(\eta_\perp,\mu)$}
    \includegraphics[height=\columnwidth, width=6cm, angle=-90]
    {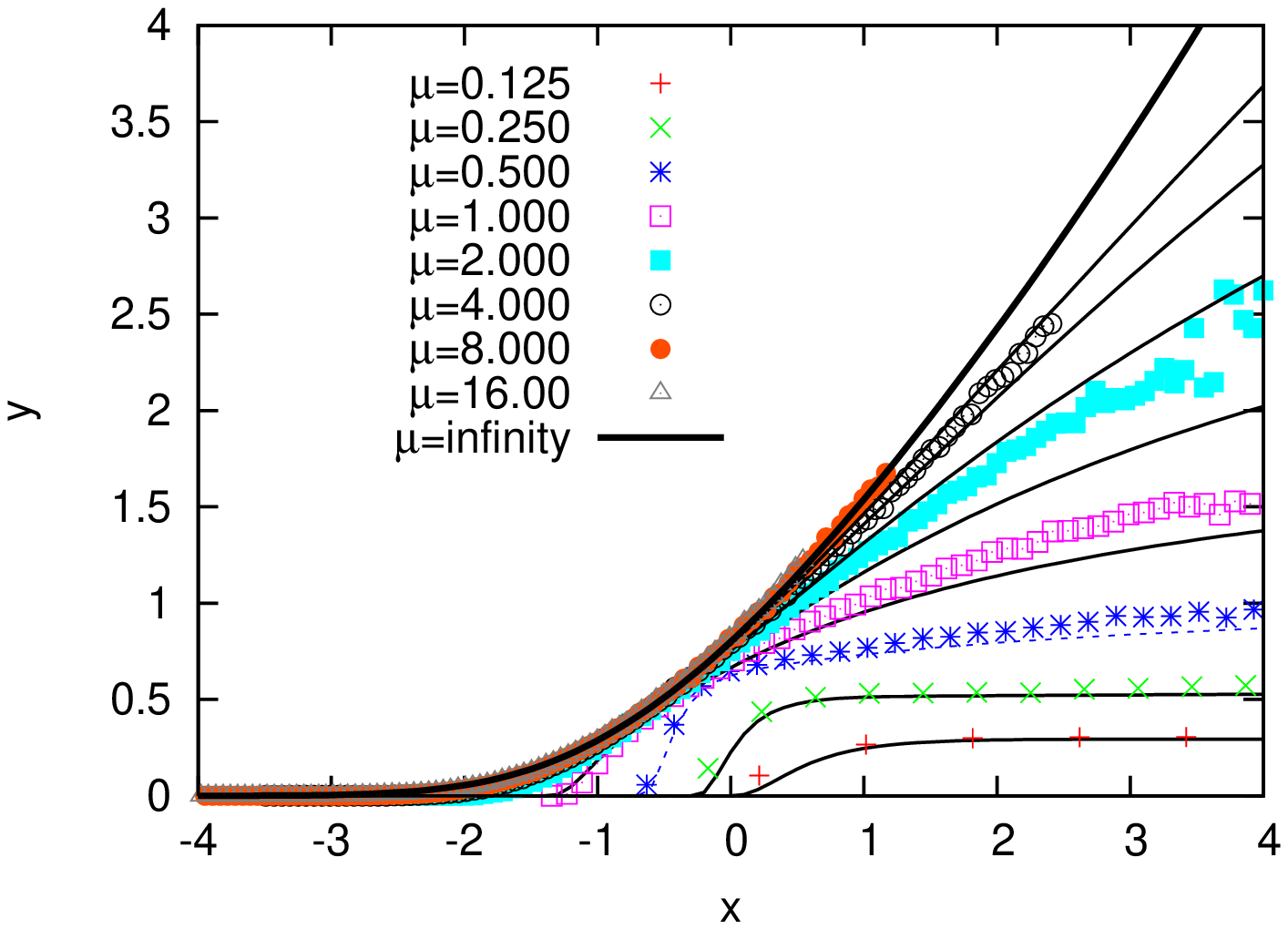}
    \rotatebox{-90}{\includegraphics[height=\columnwidth]
      {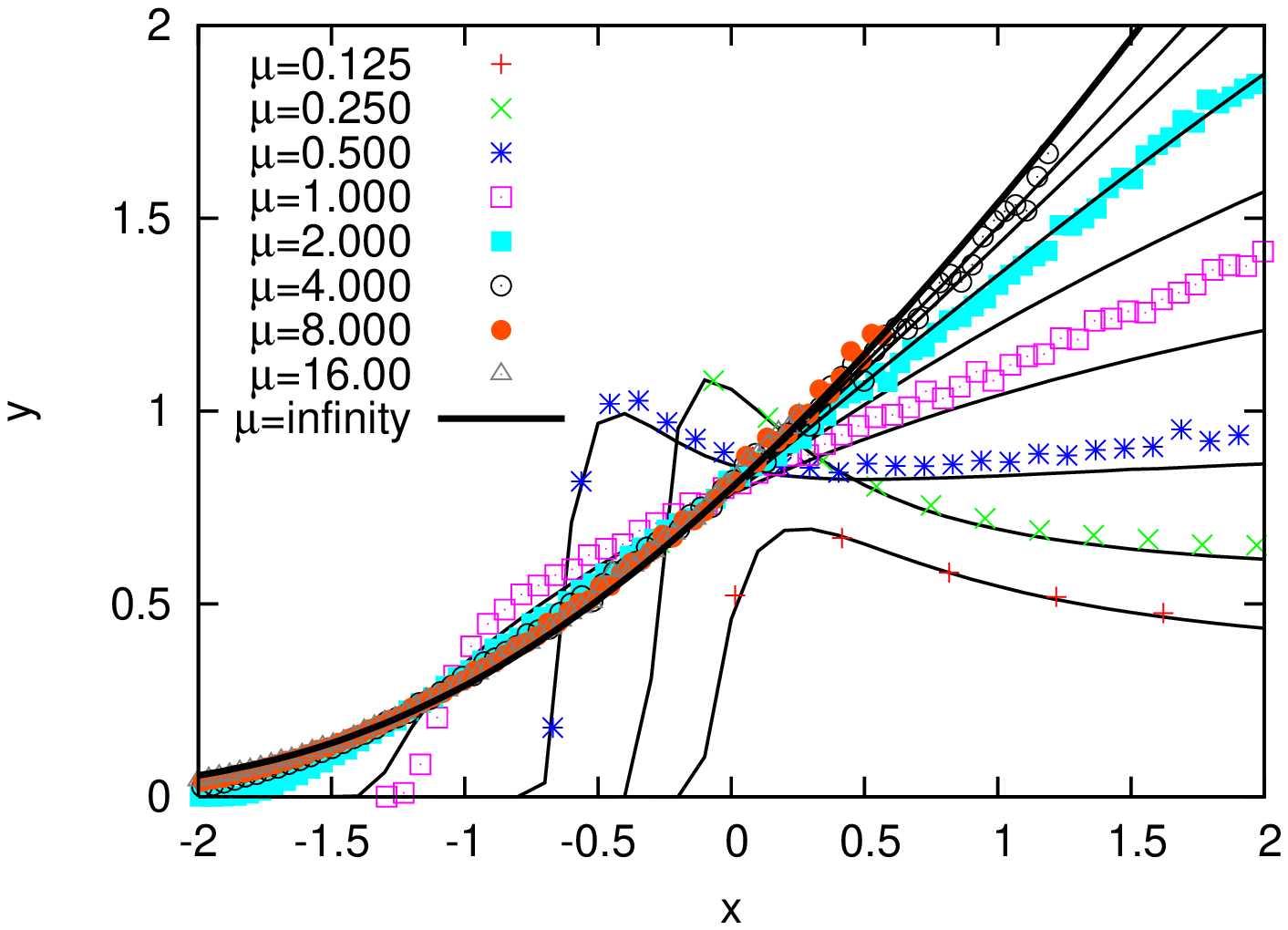}}
    \caption{Scaling function $\bar f(\eta_\perp, \mu)$ in $d=3$
      (top) and $d=2$ (bottom) for a series of values for $\mu$ (solid
      lines). For large $\mu$, the scaling function $\bar
      f(\eta_\perp, \mu)$ asymptotically converges to $\bar
      f_\perp(\eta_\perp)$ obtained within a factorization
      approximation. The MC data indicated by different symbols in the
      graphs are given for a fixed stiffness parameter $\varepsilon =
      0.1$.}
    \label{fig:fplot_perp}
  \end{center}
\end{figure}
Like in the previous scaling plot the force should vanish for $\delta
\eta_\parallel <0$, which in terms of $\eta_\perp$ reads $\eta_\perp <
- \frac32 \mu$. Again, there is a marked difference between 2d and 3d
results; see Fig.\ref{fig:fplot_perp}. We also observe that the
scaling function $\bar f(\eta_\perp, \mu)$ asymptotically approaches a
limiting curve for $\mu \to \infty$, which for a fixed value of
$\varepsilon$ corresponds to $\vartheta\to\pi/2$. It turns out, as we
will show now, that this limiting behavior can well be explained
within a factorization approximation $P(x,z) \approx P_\parallel(z)
P_\perp(x)$.  Then, ${\cal Z}(\zeta, \vartheta)$ simplifies to
\begin{eqnarray}
  {\cal Z} (\zeta, \vartheta) = \int d  z P_\parallel(z)
  {\cal Z}_\perp(\zeta \sin^{-1}\vartheta - z \cot\vartheta) \,.
\end{eqnarray}
where
\begin{eqnarray}
  {\cal Z}_\perp(x) = \int_{-\infty}^x d  x' P_\perp(x') 
\end{eqnarray}
is the restricted partition sum for the transverse fluctuations.  The
longitudinal distribution function $P_\parallel(z)$ is, for small
$L/\lp$, strongly peaked at $z \approx L$ with a characteristic width
of $\lpar$, and ${\cal Z}_\perp$ varies on the scale $\lperp$. Then,
for $ \mu \gg 1 $, the width of the longitudinal distribution function
is much smaller than the transverse restricted partition sum, such
that the integration over $P_\parallel$ can be approximated by ${\cal
  Z}(\zeta,\vartheta) \approx {\cal Z}_\perp \left( [\zeta - L
  \cos\vartheta] \sin^{-1}\vartheta \right)$ which upon using that the
transverse distribution function is a simple Gaussian,
Eq.(\ref{eq:transverse_tip_distribution}), results is
\begin{eqnarray}
{\cal Z} (\zeta,\vartheta) 
\approx \frac12 \erfc \frac{\eta_\perp}{\sqrt{2}} 
=: \bar {\cal Z}_\perp (\eta_\perp)
\end{eqnarray}
This approximation fails when $\mu \approx 1$, which defines an angle
\begin{eqnarray}
  \vartheta_c = \arctan (\lpar/\lperp) \approx \sqrt{3 L/\lp} 
\end{eqnarray}
well above which the factorization approximation is valid. The entropic
force is then
\begin{eqnarray}
  f(\zeta, \vartheta) = \frac{\kT}{\lperp\sin\vartheta}
                     \, \bar f_\perp(\eta_\perp) \, , 
\label{eqn:f_perp}
\end{eqnarray}
where
\begin{eqnarray}
   \bar f_\perp(\eta_\perp)
   = - \frac{\bar {\cal Z}_\perp'(\eta_\perp)}
            {\bar {\cal Z}_\perp(\eta_\perp)}
   = \sqrt{\frac{2}{\pi}} 
       \frac{\e^{-\eta_\perp^2/2}}{\erfc(\eta_\perp/\sqrt{2})}\,.
\label{eqn:tilde_f_perp}
\end{eqnarray}
This result for the scaling function of the entropic force is
indicated as the thick solid line in Fig.\ref{fig:fplot_perp}.  It
becomes exact in the limit $\vartheta = \pi/2$, where starting from Eq.
(\ref{eqn:Z_general_def}) one can integrate out the longitudinal
coordinate to end up with
\begin{eqnarray}
  {\cal Z}(\zeta,\frac{\pi}{2}) 
  = \frac12 \erfc \left( \frac{-\zeta}{\sqrt{2} \lperp} \right) \, .
\end{eqnarray}
Finally, for large $\zeta$, one recovers the linear response result
$f(\zeta, \pi/2) = 3 \kappa \zeta/L^3$.

If we compare the results of the factorization approximation for
$\vartheta > \vartheta_c$, \eqref{eqn:f_perp} and \eqref{eqn:tilde_f_perp},
to Eq.(2) and Eq.(5) of Ref.\cite{mogilner-oster:96}, one realizes
that they are almost identical up to the minor difference that
Mogilner and Oster define their $\kappa_0$ to be $4 \kappa/L^3$ where
it actually should be $3\kappa/L^3$. The factor $4$ in
Ref.\cite{mogilner-oster:96} instead of the correct value $3$ is the
result of assuming that the minimal energy configuration of a thin rod
bent by application of a force to its non-grafted end has constant
radius of curvature for small deflections, which is not the case. In
fact, the boundary condition of the mechanical problem forces the
curvature to vanish at the non-grafted end. In
Ref.\cite{mogilner-oster:96} the entropic force was calculated by
taking into account transverse fluctuations of the grafted polymer
only and completely disregarding any stored length fluctuations. Here,
the factorization approximation, which treats longitudinal and
transverse fluctuations as independent, gives the same result for
inclination angles $\vartheta > \vartheta_c$. The reason behind the validity
of the asymptotic results, \eqref{eqn:f_perp} and
\eqref{eqn:tilde_f_perp}, is that the tip distribution function is
much narrower in the longitudional than the transverse direction for
$\vartheta \gg \vartheta_c \sim \sqrt{L/\lp}$. Hence the range of validity
of the factorization approximation becomes larger as the polymers
become stiffer. Of course, the analysis in
Ref.\cite{mogilner-oster:96} has to fail for small inclination angles
since it does not account for stored length fluctuations at all. This
is seen most dramatically for $\vartheta = 0$, where such an
approximation would give no force at all in contrast to what we find
in Section \ref{sec:orthogonal}.

\subsection{Entropic forces: explicit results}
\label{sec:entropic_forces_explicit}

The analysis in the previous section gives the full scaling picture
for the entropic forces as a function of the scaling variables
$\eta_\parallel$ and $\eta_\perp$. Here we discuss our findings in
terms of the actual distance of the grafted end to the wall $\zeta$,
the inclination angle $\vartheta$, and the stiffness parameter
$\varepsilon=L/\lp$, which may be more convenient for actual
applications. Of course, the disatvantage of such a representation is
that we now have to give the results for particular values of the
stiffness parameter. In this section we also restrict ourselves to the
discussion of filaments which are allowed to fluctuated in 3d.

\begin{figure}[htbp]
  \begin{center}
    \psfrag{x}{\hspace{0cm}$\zeta / L$} \psfrag{y}{$f/f_c$}
    \psfrag{q}{$\vartheta$} 
    \psfrag{u}{$\vartheta=89^{\circ}$}
    \psfrag{w}{$\vartheta=17^{\circ}$}
    \rotatebox{-90}{\includegraphics[height=\columnwidth]
      {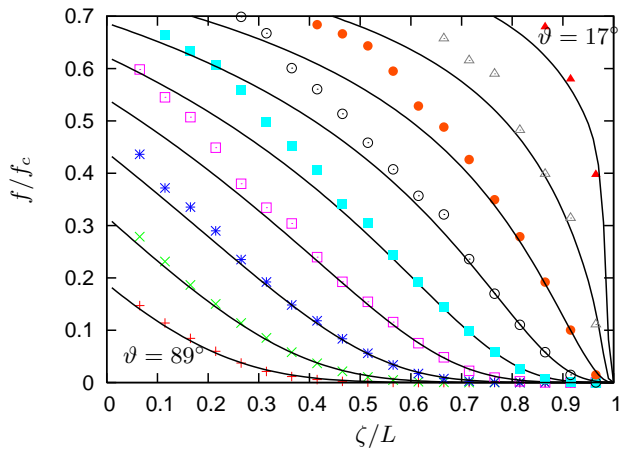}}
    \caption{Analytical and MC simulation results for the entropic 
      force $f/f_c$ as a function of the distance of the grafted end
      from the wall $\zeta/L$ for a series of inclination angles
      $\vartheta = 17,\cdots, 89$ with steps 9 (in degree).}
    \label{fig:ForceasaFunctionofZeta}
  \end{center}
\end{figure}

\begin{figure}[htbp]
  \begin{center}
    \psfrag{x}{\hspace{0cm}$\vartheta$} \psfrag{y}{\hspace{0cm}$f/f_c$}
    \rotatebox{-90}{\includegraphics[height=\columnwidth]
      {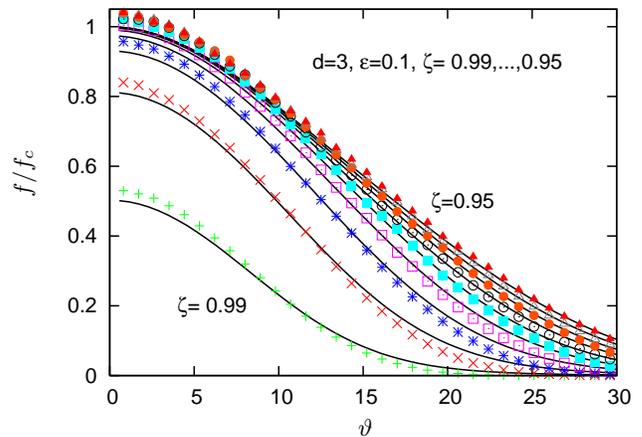}}
    \caption{Analytical and MC simulation results for the entropic 
      force $f/f_c$ as a function of the inclination angle $\vartheta$
      (in degrees) for a series of distances to the wall $\zeta /L =
      0.99, 0.985, \cdots, 0.95$.}
    \label{fig:ForceasaFunctionofThetawithoutMC}
  \end{center}
\end{figure}

In Figs.\ref{fig:ForceasaFunctionofZeta} and
\ref{fig:ForceasaFunctionofThetawithoutMC} the force $f$ in units of
the Euler buckling force $f_c$ is shown as a function of $\zeta$ (in
units of the total filament length $L$) for a series of values of
$\vartheta$ and vice versa; the stiffness parameter has been taken as
$\varepsilon = 0.1$.
Recall that the angle $\vartheta = 0$ corresponds to a wall perpendicular
to the orientation of the grafted end of the polymer, which has been
discussed in detail in Section \ref{sec:orthogonal}. Upon increasing
the inclination angle $\vartheta$ the entropic force decreases for all
given values of $\zeta$. This is to be expected since the wall then
cuts off less from the probability cloud of the polymer tip. For the
same reason the forces also decrease with increasing $\zeta$ for a
given value of $\vartheta$. The analytical results (solid lines) agree
well with the MC data for not too small values of $\zeta$. The
deviations grow larger upon decreasing the distance between the wall
and the grafted end.  Then non-linear effects not taken into account
by our weakly bending approximation set in.

\begin{figure}[htbp]
  \begin{center}
    \psfrag{x}{\hspace{0cm}$\zeta/L$} \psfrag{y}{$f/f_c$}
    \psfrag{q}{$\vartheta$}     
    \psfrag{u}{$\vartheta=89^{\circ}$}
    \psfrag{v}{$\vartheta=17^{\circ}$}
    \rotatebox{-90}{\includegraphics[height=\columnwidth]
      {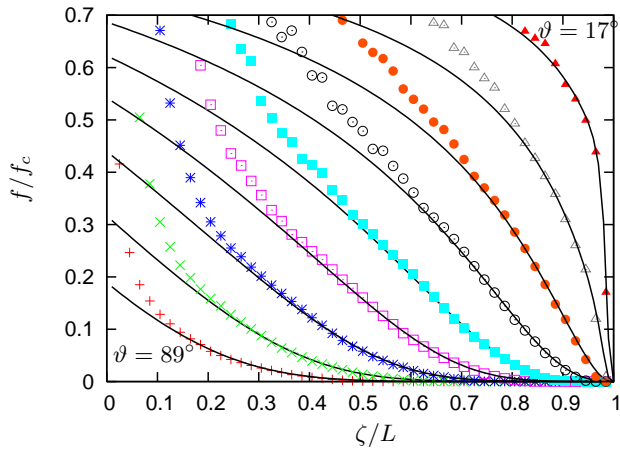}}
    \caption{Comparison of the analytical results for the entropic force 
      as a function of $\zeta / L$ (solid lines) for a series of
      values for $\vartheta = 17,\cdots, 89$ with steps 9 (in degree) indicated
      in the graph with MC simulations (symbols as indicated in the
      graph), which take into account the constraints of the wall on
      the body of the polymer; $\varepsilon = 0.1$. The deviations are
      most pronounced for small values of $\zeta$ and inclination
      angles $\vartheta$ close to $\pi/2$.}
    \label{fig:ForceasaFunctionofTheta_Contour_MC}
  \end{center}
\end{figure}

In the limit as the inclination angle approaches $\pi /2$ it is
certainly no longer justified to calculate the entropic force by
assuming that only the polymer tip is not allowed to penetrate the
membrane. Then, one has to take into account the fact that also the
body of the polymer is constrained by the presence of the wall.  Since
this reduces the number of allowed polymer configurations even further
this effect is expected to lead to an enhancement of the entropic
force. Indeed this is the case, as one may infer from
Fig.\ref{fig:ForceasaFunctionofTheta_Contour_MC}, where we show a
comparison with MC simulation accounting for these constraints. One
also notes that the enhancement of the entropic forces becomes largest
as $\vartheta \to \pi/2$ and the distance between the wall and the
grafted end becomes small; a full account of this effect will appear
in Ref.\cite{gholami_frey:in_preparation} .

\begin{figure}[htbp]
  \begin{center}
    \psfrag{x}{\hspace{0cm}$\zeta$} \psfrag{y}{$f/f_c$} \psfrag{q
    }{results of Ref[2], (Eq. (\ref{eqn:tilde_f_perp}))} \psfrag{h
    }{our results, $\vartheta=80^\circ,..,30^\circ$}
    \psfrag{t}{$\vartheta=30^\circ$} \psfrag{z}{$\vartheta=80^\circ$}
    \rotatebox{-90}{\includegraphics[height=\columnwidth]
      {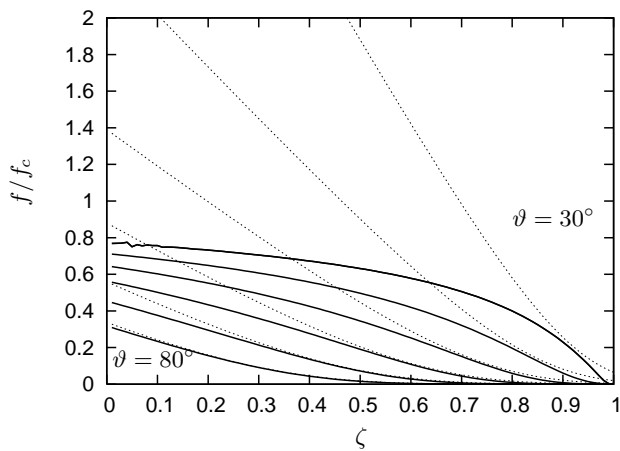}}
    \caption{Comparison of the full result for the entropic force as
      a function of $\zeta$ (full lines) with the results obtained
      from the factorization approximation (dashed lines) for a
      stiffness parameter $\varepsilon = 0.1$ and a series of inclination
      angles $\vartheta=$ $30^\circ, 40^\circ, \cdots, 80^\circ$. The
      range of validity of the factorization approximation broadens as
      one increases the inclination angle $\vartheta$; it is invalid for
      $\vartheta$ smaller than $\vartheta_c \approx 30^\circ$.}
    \label{fig:fplot_Force3D_Com_Mogilner}
  \end{center}
\end{figure}

Finally, we would like to compare our full results with the
factorization approximation discussed in the previous section,
Eq.(\ref{eqn:tilde_f_perp}), which when corrected for some minor
factor is identical to the results given in
Ref.\cite{mogilner-oster:96}. The comparison is given in
Fig.\ref{fig:fplot_Force3D_Com_Mogilner} for a stiffness parameter
$\varepsilon = 0.1$.
In the limit of large inclination angles well above $\vartheta_c \approx
30^\circ$, there is excellent agreement between the factorization
approximation and the full results for not too small values of
$\zeta$. As one approaches $\vartheta_c$ the range of validity of the
factorization approximation shrinks and finally it becomes invalid for 
$\vartheta < \vartheta_c$. 

To illustrate the applicability of the factorization approximation let
us take some examples. For the cytoskeletal filament F-actin with a
contour length $100$~nm and persistence length $\lp=15 \mu$m, the
stiffness parameter becomes $\varepsilon=0.006$ which gives $\vartheta_c
\approx 7.6^\circ$. Upon increasing the stiffness parameter to
$\varepsilon=0.1$, which amounts to changing the contour length to a
value of $L = 1.6\mu$m, the critical angle $\vartheta_c$ increases to
$28.7^\circ$. 

\section{Summary and conclusions}

In summary, we have presented analytical calculations and extensive
Monte Carlo simulations for the entropic force $f$ exerted by a
grafted polymer on a rigid obstacle (wall). The scale for the
magnitude of the entropic force is given by the {\em Euler buckling
  force} $f_c \propto \kT \lp / L^2$. The {\em stiffness parameter}
$\varepsilon = L / \lp$ discerns the two universal regimes of a
Gaussian chain ($\varepsilon \gg 1$) and a semiflexible chain
($\varepsilon \ll 1$).  In this manuscript we have mainly focused on
the {\em stiff limit}, where analytical calculations using a
weakly bending rod approximation are possible. In comparing our
results with Monte Carlo simulations we have found that the range of
applicability of the results obtained in the stiff limit extend
to stiffness parameters as large as $\varepsilon = 0.1$. Qualitatively
the asymptotic results remain valid even up to $\varepsilon = 1$. 

For the simplest possible geometry, where the polymer is perpendicular
to the wall, located at a distance $\zeta$ from the grafted end, our
analytical calculations show that the entropic force obeys a scaling
law in the stiff limit
\begin{eqnarray}
  f_\parallel (\zeta, L, \lp) = f_c \tilde f_\parallel (\etatil)
\end{eqnarray}
with the scaling variable $\etatil = (L-\zeta)/\lpar$ measuring the
minimal compression of the filament in units of the longitudinal width
of the tip distribution function $\lpar = L^2/\lp$, and $f_c$ the
Euler buckling force of a classical beam. For small values of the
scaling variable we have derived a simple analytical expression,
\eqref{eqn:f_form3},
\begin{eqnarray}
  \tilde f_\parallel^<(\etatil)
  = \frac{4}{\pi^{5/2}} \; 
    \frac{\exp [ -1/4 \etatil]}
         {\etatil^{3/2} 
          \left[ 1 - 2 \erfc(1/2\sqrt{\etatil}) \right]}
\end{eqnarray}
and a corresponding formula in 2d, \eqref{eqn:forceform2_2d}, which
describe the full scaling function to a high numerical accuracy for
$\etatil \leq 0.2$. For $\etatil \geq 0.2$ there are equally simple
expressions, as for example \eqref{eqn:f_form4} for 3d. We expect
these formulas to be useful for molecular models of cell motility. The
shape of the scaling function shows dramatic differences between 2d
and 3d, which are of geometric origin. In 3d the entropic forces
always stay below the Euler buckling force. In contrast, in 2d it is
larger than the mechanical limit for most of the parameter space and
exhibits a pronounced maximum at small values of the scaling variable
$\etatil$ before it steeply drops to zero as $\zeta \to L$. 

Extensive Monte Carlo simulations confirm these analytical results and
show that their range of applicability is $\varepsilon \leq 0.1$. For
larger values of the stiffness parameter there are clear deviations
from the stiff scaling limit, which become qualitative for
$\epsilon \geq 1$. Features of the stiff limit, such as the
maximum in the entropic force, are visible even for $\varepsilon$ as
large as $4$. 

For a polymer inclined at an angle $\vartheta$ with respect to the wall
also the transverse width $\lperp = \sqrt{L^3/3\lp}$ of the tip
distribution function plays a significant role; note that the ratio
$\lpar / \lperp = \sqrt{3 \varepsilon}$. The entropic force can now be
written in the scaling form
\begin{eqnarray}
  f(\zeta, \vartheta; L, \lp) 
  =  f_c (\vartheta) 
    \tilde f (\eta_\parallel, \eta_\perp) \, , 
\end{eqnarray}
where $\eta_\perp = ( L\cos\vartheta-\zeta )/( L_\perp \sin\vartheta )$, 
$\eta_\parallel = ( L\cos\vartheta-\zeta )/( L_\parallel \cos\vartheta )$ and
$f_c (\vartheta) = f_c / \cos\vartheta$. It turned out that a proper choice
of scaling variables are $\mu = \eta_\parallel / \eta_\perp =
(\lperp/\lpar) \tan\vartheta $ and $\eta_\parallel$ or $\eta_\perp$
depending on whether the inclination angle is smaller or larger than a
characteristic angle $\tan \vartheta_c = \lpar / \lperp$, i.e. $\mu_c=1$.
Upon increasing the inclination parameter $\mu$ the shape of the
scaling function changes from a step-function-like form at $\mu = 0$
to a purely convex shape as $\mu \to \infty$.The limit $\mu \to
\infty$ either corresponds to $\vartheta \to \pi / 2$ or for a fixed
$\vartheta \neq 0$ to the stiff limit $\varepsilon \to 0$. For 2d, in
addition, the maximum vanishes at $\mu \approx 0.6$.

In the limit of inclination angles which are much larger than the
characteristic angle $\vartheta_c$, we have found that an approximation,
\eqref{eqn:f_perp} and \eqref{eqn:tilde_f_perp}, based on factorizing
the joint probability distribution of the polymer tip gives an
excellent asymptotic representation of the full analytical results:
\begin{eqnarray}
  f(\zeta, \vartheta) 
     = \frac{\kT}{\lperp\sin\vartheta}
       \sqrt{\frac{2}{\pi}} 
       \frac{\e^{-\eta_\perp^2/2}}{\erfc(\eta_\perp/\sqrt{2})}\,.
\end{eqnarray}
It is simpler than the full scaling form since it only depends on one
scaling variable. Up to minor factors this asymptotic formula for the
entropic force is mathematically identical to the results found in
Ref.\cite{mogilner-oster:96}, which was derived upon assuming that the
tip of the polymer fluctuates perpendicular to its contour only. 
Since $\tan \vartheta_c \propto \sqrt{\varepsilon}$ the range of
applicability of this results grows with increasing stiffness
parameter. For example, $\vartheta_c$ equals approximately $30^\circ$ and
$10^\circ$ for stiffness parameter $\varepsilon$ equal to $0.1$ and
$0.006$, respectively. For $\vartheta \leq \vartheta_c$ the factorization
approximation fails completely, since it gives an incorrect
description of the longitudinal stored length fluctuations. Then, a
full analysis in terms of a two parameter scaling function is necessary.

\acknowledgments It is a pleasure to acknowledge helpful discussions
with Panayotis Benetatos, Martin Falcke, Thomas Franosch, Klaus Kroy,
Gianluca Lattanzi and Frederik Wagner. This work was supported by the
Deutsche Forschungsgemeinschaft (Research Training Group 268 on
``Dynamics and evolution of cellular and macromolecular processes'').

\begin{appendix}

\section{Inverse Laplace transform of the moment generating function}
\label{app:inverse_laplace}

In this appendix we collect our calculations of the inverse Laplace
transform of the moment generating functions. This will give as two
sets of series representations, which show good convergence properties
either close to full stretching or for strong compression of the
filament.

\subsection{Series representation of the 3d tip distribution function
  for large stored length}
\label{app:inverse_laplace3d}

Starting from the moment generating function ${\cal P}_\parallel (f)$
one can calculate the distribution function $P_\parallel (z)$ by an
inverse Laplace, i.e. an integral along the imaginary axis,
\begin{eqnarray}
\label{eqn:BackLaplaceTransform}
 P_\parallel (z) 
 = \int_{-\i \infty}^{+\i \infty} 
   \frac{df}{2 \pi \i}
   \, \e^{f (L-z)} \, {\cal P}_\parallel (f) \, .
\end{eqnarray}
Since the moment generating function
\begin{eqnarray}
 {\cal P}_\parallel (f) 
  = \prod_{k=1}^\infty
      \left( 1 + \frac{4 f L^2}{\lp (2k-1)^2 \pi^2}
      \right)^{-1}
\end{eqnarray}
has poles at $f_k = -\lambda_k^2 \lp/L^2$ with $k=1,2,3,...$ only
along the negative real axis, standard residuum calculus gives
\begin{widetext}
\begin{eqnarray}
  {\cal P}_\parallel (f)
  = \sum_{k=1}^{\infty}
    \exp \left[ -(L-z) \, \lambda_k^2 \, \frac{\lp}{L^2} \right]
    \times \prod_{l \neq k} 
    \left( 1 - \frac{(2k-1)^2}{(2l-1)^2} \right)^{-1}
    \left( \frac{L^2}{\lp \lambda_k^2}\right)^{-1}
\end{eqnarray}
Using $\prod_{k=1}^\infty \left( 1 - \frac{x^2}{(2k-1)^2} \right)
  = \cos \left(\frac{\pi}{2} x \right)$ \cite{Stegun},
the product term can be written as
\begin{eqnarray}
  \prod_{l \neq k} \left( 1- \frac{(2k-1)^2}{(2l-1)^2} \right)^{-1}
  &=& \lim_{k' \to k} \left( 1 - \frac{(2k'-1)^2}{(2k-1)^2} \right)
      \prod_l \left( {1-\frac{(2k'-1)^2}{(2l-1)^2}} \right)^{-1}
      \nonumber \\
  &=& \lim_{k' \to k} \left(1 - \frac{(2k'-1)^2}{(2k-1)^2} \right)
      \cos^{-1} \left( \frac{\pi}{2} (2 k'-1) \right)
      \nonumber \\   
   &=& 2 \frac{(-1)^{k+1}}{\pi} \frac{2}{2k -1}
   = 2 (-1)^{k+1} \, \frac{1}{\lambda_k} \, .
\end{eqnarray}
Hence we find
\begin{eqnarray}
  P_\parallel(z) = 2 \lpar^{-1} \sum_{k=1}^\infty (-1)^{k+1}
  \lambda_k
  \exp \left[- \lambda_k^2 (L-z)/\lpar \right]
\end{eqnarray}
with the characteristic longitudinal length scale $\lpar = L^2 / \lp$.

\subsection{Series representation for the tip distribution function
  close to full stretching: general $d$}
\label{app:inverse_laplace2d}

We begin the analysis with the two-dimensional case, where 
\begin{eqnarray}
{\cal P}_\parallel (f) 
 = \prod_{k=1}^\infty 
   \left(1 + \frac{f \lpar}{\lambda_k^2} \right)^{-1/2} 
 = \sqrt{\frac{1}{\cosh \sqrt{f \lpar}}} \, .
\end{eqnarray}
For the derivation of our first series representation we start from
the product formula for the moment generating function. In this
representation one has branch cuts on the negative real axis at $
\tilde f = f \lpar = - \lambda_k^2$ for $k \in \mathbb{N}$. We now
deform the contour in the complex plane such that we enclose the
negative real axis. Then
\begin{eqnarray}
 \tilde P_\parallel (\rhotil) 
 &=& \int_{-\i \infty}^{+\i \infty} \frac{d \tilde f}{2 \pi \i} \, 
   \e^{\tilde f \rhotil} \, 
   \tilde {\cal P}_\parallel (\tilde f) 
\nonumber \\
 &=& \int_{-\infty}^{0} \frac{d \tilde f}{2 \pi \i} \, 
   \e^{\tilde f \rhotil} \, 
   \tilde {\cal P}_\parallel (\tilde f - \i\epsilon)
 + \int_{0}^{-\infty} \frac{d \tilde f}{2 \pi \i} \, 
   \e^{\tilde f \rhotil} \, 
   \tilde {\cal P}_\parallel(\tilde f +\i\epsilon)
\nonumber \\
 &=&  \int_{0}^{\infty} \frac{d \tilde f}{2 \pi \i} 
     \, \e^{- \tilde f \rhotil} \, 
     \left[ \tilde {\cal P}_\parallel (- \tilde f -\i\epsilon) - 
            \tilde {\cal P}_\parallel (- \tilde f +\i\epsilon)
     \right] 
\end{eqnarray}
where $\epsilon \to 0$. To proceed we need to evaluate the product
formula on the negative real axis. We find for $x \in
[2k+1,2k+3]\frac{\pi}{2}$
\begin{eqnarray}
 \lim_{\epsilon \to 0}\prod_{l=1}^{\infty} 
 \, \sqrt{1 - \frac{x^2 \mp \i \epsilon}{\lambda_l^2}}
 = (\mp \i)^k \frac{1}{\sqrt{|\cos x|}} 
\end{eqnarray}
Upon substituting $y^2 = \tilde f$ this finally results in the series
expansion
\begin{eqnarray}
 \tilde P_\parallel (\rhotil) 
  = \frac{2}{\pi} 
    \sum_{n=0}^{\infty} (-1)^n 
    \int_{\lambda_{2n+1}}^{\lambda_{2n+2}} dy \, 
    \frac{y \, \e^{-y^2 \rhotil}} {\sqrt{|\cos y|}} \, .
\end{eqnarray}
For large values of $\rhotil$, corresponding to a significant
compression of the polymer, the integral is dominated by the
contribution from the interval $[\pi/2,3\pi/2]$, such that the leading
factor will be proportional to $\exp(-\pi^2\rhotil/4)$. In order to
evaluate $\tilde P_{\parallel}(\rhotil)$ further, we may average $y
e^{-y^2 \tilde \rho}$ over the interval and approximate
the integral as
\begin{eqnarray}
 \int_{\lambda_{2 n+1}}^{\lambda_{2n+2}} dy 
 \frac{y\, \e^{-y^2 \rhotil}}{\sqrt{|\cos y|}}
 \approx 
   \frac{1}{5} \sum_{m=4}^{8}
    \lambda_{2n+\frac{m}{4}} \, 
    \exp \left[-\lambda_{2n+\frac{m}{4}}^2 \, \rhotil \right]
   \int_{\frac{\pi}{2}}^{\frac{3\pi}{2}}\frac{dy}{\sqrt{|\cos y|}}
\end{eqnarray}
such that we finally get
\begin{eqnarray}
 \tilde P_\parallel (\rhotil) 
 \approx \frac{1}{\mathcal{N}}
 \sum_{n=0}^{\infty}(-1)^n
 \sum_{m=4}^8
    \lambda_{2n+\frac{m}{4}}
    \exp \left[ -\lambda_{2n+\frac{m}{4}}^2 \, \rhotil \right] \, , 
\label{eqn_pofz2dSecondForm}
\end{eqnarray}
where 
\begin{eqnarray}
 \mathcal{N}^{-1} 
 = \frac{2}{5 \pi} 
   \int_{\frac{\pi}{2}}^{\frac{3\pi}{2}}\frac{dy}{\sqrt{|\cos y|}} 
 \approx 0.67 \, .
\end{eqnarray}
\end{widetext}

Next we drive a series representation suitable for small values of
$\rhotil$. We use that for $f\in\mathbb{R}_+$ one has \cite{Hansen}
\begin{eqnarray}
 {\cal P}_\parallel (f) 
 = \frac{1}{\sqrt{\cosh \sqrt{f \lpar}}} \,.
\end{eqnarray}
With $\cosh (x)= \frac12 (\e^x+\e^{-x})$ and the generalized binomial
theorem, this can be expanded to give 
\begin{eqnarray}
 {\cal P}_\parallel (f) 
 = \sqrt{2} \sum_{l=0}^{\infty}
   \left(\begin{array}{c}  
           - \frac12 \\ 
           l 
         \end{array} \right)
   \e^{-(2l+1/2)\sqrt{f \lpar}} \, ,
\label{eq:series_expansion_mgf}
\end{eqnarray}
which is a holomorphic function on $\mathbb{C}\backslash
\mathbb{R}_-$. Hence by the theorem of identity from complex calculus
this formula remains valid $ \forall f \in \mathbb{C}\backslash
\mathbb{R}_-$.  Substituting $y=\sqrt{f \lpar}$ transforms
\eqref{eqn:BackLaplaceTransform} to
\begin{eqnarray}
 \tilde P_\parallel (\rhotil) 
 = \int_{-i\infty+\varepsilon}^{i\infty+\varepsilon}
   \frac{dy}{\pi \i} \, \e^{y^2 \rhotil} \, y \, 
   \tilde {\cal P}_\parallel (y^2) \, .
\end{eqnarray} 
Inserting the series representation \eqref{eq:series_expansion_mgf}
and using the integral representation 
\begin{eqnarray}
  D_1 (z) = \sqrt{2\pi}\e^{\frac{z^2}{4}}
          \int_{-i\infty+\varepsilon}^{i\infty+\varepsilon} 
          \frac{ds}{2\pi \i} \, s \, 
          \exp \left[ -zs+\frac{s^2}{2} \right]  
\end{eqnarray}        
for the parabolic cylinder function \cite{Stegun} as well as
\begin{eqnarray}
 \left( \begin{array}{c} 
          - \frac12 \\ l 
        \end{array}\right)
 = (-1)^l \frac{(2l-1)!!}{2^ll!} \, , 
\label{eq:binomial}
\end{eqnarray}
where $n!!=n(n-2)(n-4)\ldots$ yields
\begin{eqnarray}
 \tilde P_\parallel (\rhotil) 
 &=& \frac{1}{\sqrt{\pi} \rhotil}
     \sum_{l=0}^{\infty} (-1)^l \frac{(2l-1)!!}{2^ll!}
 \nonumber \\
 &\times& \exp \left[ - \frac{(l+\frac14)^2}{2\rhotil}\right]
          D_1\left[ \frac{2l+\frac12}{\sqrt{2\rhotil}}
         \right] \, .
\label{appA:A19}
\end{eqnarray}
With $D_1(x) = x \e^{-x^2/4}$ \eqref{appA:A19} becomes
\eqref{eqn_pofz2d}.
 
Finally, all the calculations are easily generalized to general
spatial dimensions $d$. One finds the series representation
\begin{eqnarray}
 \tilde P_\parallel (\rhotil) 
 &=& 2^{d/2} \frac{1}{\sqrt{2\pi}}
     \sum_{l=0}^{\infty} 
     \left(\begin{array}{c}-{\frac12(d-1)} \\ l \end{array}\right) 
 \nonumber \\
 &\times& \frac{l +\frac14(d-1)}{\rhotil^{3/2}}
     \exp \left[ -\frac{(l+\frac{d-1}{4})^2}{\rhotil}\right]
\end{eqnarray}
which is the fast converging for small $\rhotil$.

\section{Saddle point approximation}
\label{app:saddle_point}

Starting from \eqref{eqn:BackLaplaceTransform} and introducing
$\tilde f=f \lpar$ gives
\begin{eqnarray}
\label{Eqn:(Saddle1)}
  P_\parallel (z) 
  &=& \int_{-\i \infty}^{+\i \infty} \frac{df}{2 \pi \i} \, 
  \e^{f \lpar \rhotil} \, \cosh^{-1} \sqrt{f \lpar} 
 \nonumber \\
 &=&\lpar^{-1}\int_{-\i \infty}^{+\i \infty} \frac{d \tilde f}{2 \pi \i} \,
 \frac{2 \e^{\tilde f \rhotil}}
      {\e^{\sqrt{\tilde f}}+\e^{-\sqrt{\tilde f}}} \, .
\end{eqnarray}
We are interested to the asymptotic result of the integral close to
full stretching $\rhotil \rightarrow 0$. Upon substituting $\tilde f =
\xi/\rhotil^2$ one finds
\begin{eqnarray}
\label{Eqn:(Saddle2)}
 P_\parallel (z) 
 = \frac{2}{\rhotil^2 \lpar}
   \int_{-\i \infty}^{+\i \infty} d \xi \, 
   \frac{\exp[f(\xi)/\rhotil]}{1+\exp[-2\sqrt{\xi}/\rhotil]}
\end{eqnarray}
where $f(\xi)=\xi-\sqrt \xi$. Since the function $f(\xi)$ has a global
maximum at $\xi_{0}=0.25$ the main contribution to the integral in the
limit $1/\rhotil \rightarrow \infty$ comes from the integration along
the curve of steepest descent which passes through $\xi_0$. We need to
find this curve such that $\Im[f(\xi)] = \text{constant} =
\Im[f(\xi_0)] = 0$. We write $\sqrt \xi=\sqrt a(1+\i s)$ in terms of
the curve parameter $s$. Then the condition $\Im[f(\xi_0)]=0$ gives
$a=1/4$, and the curve of steepest descent is given in terms of
$\Re[\xi]=\frac{1}{4}(1-s^2)$ and $\Im[\xi]=2as$, which is a parabola
parameterized by $s$. The saddle point approximation amounts to a
contour integral along this parabola, where $f(\xi) = -(1+s^2)/4$,
such that
\begin{eqnarray}
\label{Eqn:(Saddle3)}
 P_\parallel (z) 
 = \frac{1}{\lpar \rhotil^2}
    \int_{-\infty}^{\infty} \frac{ds}{2 \pi} \, (1 + \i s) \, 
    \frac{\e^{- (1+s^2)/4\rhotil}}{1+\e^{-(1+\i s)/\rhotil}} \, . 
\end{eqnarray} 
To the leading order in $\rhotil$ we get
\begin{eqnarray}
\label{Eqn:(Saddle4)}
 P_\parallel (z) 
 &=&\frac{\exp[-{1}/{4\rhotil}]}{\rhotil^2 \lpar}
    \int_{-\infty}^{+\infty} \frac{ds}{2\pi} \, 
    \exp \left[ \frac{-s^2}{4\rhotil} \right]
 \nonumber\\
 &=&\frac{1}{\sqrt{\pi\rhotil^3} \lpar}
    \exp \left[-\frac{1}{4\rhotil} \right]
\end{eqnarray}

In the two dimensional case (2d), using the same strategy and substituting
$\tilde f = \xi/\rhotil^{4/3}$ gives
\begin{eqnarray}
  P_\parallel (z) 
  = \frac{1}{\sqrt{8\pi\rhotil^3}\lpar} 
    \exp \left[-\frac{1}{16\rhotil} \right] \, .
\end{eqnarray}
  
\section{Jacobi Transformation of the restricted partition sum
  ${\cal Z}_\parallel (\zeta)$}
\label{app:jacobi_transform_parallel}

To unclutter the formulas in this section, we use the generic argument
$x$ with $x \equiv \eta_\parallel$.  ${\cal Z}_\parallel(x)$ can be
written as
\begin{eqnarray}
  \label{eqn:Z_form2}
  \tilde {\cal Z}_\parallel(x)
  &=& 2 \int_0^\infty d  y \sum_{k=-\infty}^\infty
      (-1)^{k+1} \delta (y - \lambda_k) \frac{1}{y}
      \e^{-\lp y^2 x}
  \nonumber \\
  &=& 2\int_0^\infty d  y \, \tilde\delta(y)
      \frac{1}{y} \e^{-\lp y^2 x}
\end{eqnarray}
where we defined
\begin{eqnarray}
  \tilde\delta(y) := \sum_{k=-\infty}^\infty (-1)^{k+1}
  \delta(\lambda_k-y) \,.
\end{eqnarray}
Since $\tilde\delta(y)$ is odd in $y$ and has periodicity $2\pi$, we
can expand it into a Fourier-sine-series:
\begin{eqnarray}
  \tilde\delta(y) = \sum_{l=1}^\infty d_l \sin(l y)
\end{eqnarray}
where
\begin{eqnarray}
  d_l
  &=& \frac{2}{\pi} \int_0^\pi d  y \, \tilde\delta(y) \sin(ly)
  \nonumber \\
  &=& \frac{2}{\pi} \sin(l \pi/2) \nonumber \\
  &=& \frac{2}{\pi}
  \begin{cases}
    0 & \text{if $l$ is even} \\
    (-1)^{\frac{l-1}{2}} & \text{if $l$ is odd}
  \end{cases} \,.
\end{eqnarray}
This results in
\begin{eqnarray}
  \tilde\delta(y) = \frac{2}{\pi} \sum_{l=1}^\infty (-1)^{l+1}
  \sin[(2l-1) y] \,.
\end{eqnarray}
Inserting this into \eqref{eqn:Z_form2} we find for $\tilde
{\cal Z}_\parallel(x)$
\begin{eqnarray}
  \tilde {\cal Z}_\parallel(x) = \frac{4}{\pi} \sum_{l=1}^\infty (-1)^{l+1}
  \int_0^\infty d  y \, y^{-1} \e^{-y^2 x} \sin[(2l-1) y] \,\nonumber\\.
\end{eqnarray}
The integral evaluates to \cite{Stegun} (with $\mu = 0$, $\beta = x$,
$\gamma = 2l -1$)
\begin{eqnarray}
  &&\int_0^\infty d  y \, y^{-1} \e^{-y^2 x} \sin[(2l-1) y]
  \nonumber \\
  &&=
  \frac{(2l-1) \e^{-(2l-1)^2/4 x}}{2 \sqrt{x}} \sqrt{\pi}
    {}_1F_1\left(1; \frac{3}{2}; \frac{(2l-1)^2}{4
        x}\right) \,\nonumber\\.
\end{eqnarray}
As the confluent hypergeometric function ${}_1F_1 (\alpha; \gamma; z)
\equiv \Phi(\alpha, \gamma; z)$ has the property $\Phi(\alpha; \gamma;
z) = \e^z \Phi(\gamma-\alpha, \gamma; -z)$ \cite{Stegun} we find with
\cite{Stegun}
\begin{eqnarray}
  \Phi\left(1, \frac{3}{2}; z\right)
  &=& \e^z \Phi\left(\frac{1}{2},
  \frac{3}{2}; -z\right)
  \nonumber \\
  &=& \frac{\sqrt{\pi} \e^z}{2 \sqrt{z}} \erf \sqrt{z} \,.
\end{eqnarray}
Our result for $\tilde {\cal Z}_\parallel(x)$ is thus
\begin{eqnarray}
  \tilde {\cal Z}_\parallel(x)
  = 2 \sum_{l=1}^\infty (-1)^{l+1} \erf \frac{2l-1}{2 \sqrt x}
\end{eqnarray}
This still has problems for $x \to 0$ where $\erf[(2l-1)/2 \sqrt{x}]
\to 1$. We can, however rewrite it to
\begin{eqnarray}
  \tilde {\cal Z}_\parallel(x) = 2 \sum_{l=1}^\infty (-1)^{l+1} +
  2 \sum_{l=1}^\infty (-1)^l \erfc \frac{2l-1}{2\sqrt{x}} \,.
\end{eqnarray}
All convergence problems are now isolated in the first sum. As we know
that $\tilde {\cal Z}_\parallel(0) = 1$ (compare \eqref{eqn:Z_form1}) we
assign $2 \sum_{l=1}^\infty (-1)^{l+1} = 1$ to finally find
\begin{eqnarray}
  \label{eqn:Z_form3}
  \tilde {\cal Z}_\parallel(x) = 1 + 2 \sum_{l=1}^\infty (-1)^l \erfc
  \frac{2l-1}{2\sqrt{x}}\,.
\end{eqnarray}

\section{Graft-angle-dependent force}
\label{app:TechnicalDetailsofIntegrals}
We evaluate the general expression \eqref{eqn:Z_general_def} using the
representation

\begin{eqnarray}
  \Theta(x) = \lim_{\varepsilon\to 0+} \int \frac{d
    q}{2\pi\i} \frac{\e^{\i q x}}{q - \i \varepsilon}
\end{eqnarray}
of the step function $\Theta(x)$. With \eqref{eqn:Pfull} we find

\begin{eqnarray}
  {\cal Z}(\zeta, \vartheta)
    &=& \int \frac{d q}{2\pi\i} \frac{\exp\left[\i q
    \frac{\zeta/\cos\vartheta - L}{\lpar}\right]}{q - \i \varepsilon}
    \times\nonumber \\
    &&\int d \tilde x d \etatil \e^{\i q \etatil}
    \e^{-\i q (\lperp/\lpar)
    \tan\vartheta \tilde x}
    \tilde P(\tilde x, \etatil)
    \nonumber \\
    &=& \int \frac{d q}{2\pi\i} \frac{\exp\left[\i q
    \frac{\zeta/\cos\vartheta - L}{\lpar}\right]}{q - \i
    \varepsilon} a_d(-\i q) \nonumber\\
    &&\exp\left[-( q \lperp \lpar^{-1}
    \tan\vartheta)^2 3 b(-\i q)/2\right]
    \nonumber \\
    &=& \tilde{\cal Z}\left(\frac{L - \zeta/\cos\vartheta}{\lpar},
    \frac{\lperp}{\lpar}\tan\vartheta \right)
\end{eqnarray}
where

\begin{eqnarray}
  \tilde{\cal Z}(\eta_\parallel, \mu) = - \int \frac{d  q}{2\pi\i}
  \frac{\e^{\i q }}{q + \i \varepsilon} a_d(\i q)
  \e^{- \frac{3 \mu^2 q^2 b(\i q)}{2}} \,.
\end{eqnarray}
Using the Dirac formula

\begin{eqnarray}
  \frac{1}{q + \i \varepsilon}
  = {\cal P} \frac{1}{q} - \i \pi \delta(q)  \, ,
\end{eqnarray}
$a_3(0) = 1$, $3 b(0) = 1$ and the symmetry properties of $a_3(\i q)$
and $b(\i q)$, we find

\begin{eqnarray}
  \label{eqn:Z_force_num1}
  \tilde{\cal Z}(\eta_\parallel, \mu)
  = \frac{1}{2} - 2 \int_0^\infty
    \frac{d  q}{2\pi} \frac{1}{q} \Im\left(\e^{\i q \eta_\parallel} a_3(\i q)
    \e^{-\frac{1}{2} \mu^2 q^2 3 b(\i q)}\right) \,.
\end{eqnarray}

The notation $\cal P$ denoting the principal value has been dropped as
the integrand is regular at $q = 0$.  For large $\mu$ and/or $\zeta$,
$\tilde{\cal Z}(\eta_\parallel, \mu)$ vanishes.  This means that the
integral in \eqref{eqn:Z_force_num} must approach $1/2$. Subtracting
the result of the numerically evaluating the non-vanishing integral
from $1/2$ strongly amplifies the unavoidable round-off error. We
therefore rewrite \eqref{eqn:Z_force_num1} to
\begin{eqnarray}
  \label{eqn:Z_force_num2}
  &&\tilde{\cal Z}(\eta_\parallel, \mu) = \frac{1}{2} \erfc
  \frac{\eta_\parallel}{\sqrt{2}\mu}
  \nonumber \\
  &&- 2 \int_0^\infty
  \frac{d  q}{2\pi} \frac{1}{q} 
  \Im\left[\e^{\i q \eta_\parallel} \left(a_3(\i q)
    \e^{-\frac{1}{2} \mu^2 q^2 3 b(\i q)}
  - \e^{-\frac{\mu^2 q^2}{2}}\right)\right]\nonumber\\
\end{eqnarray}
where we used the identity
\begin{eqnarray}
  \frac{1}{2} - {\cal P} \int_{-\infty}^\infty \frac{d
    q}{2 \pi \i} \frac{\e^{\i q \eta_\parallel}}{q}
  \e^{-\frac{q^2 \mu^2}{2}} = \frac{1}{2} \erfc
  \frac{\eta_\parallel}{\sqrt{2}\mu} \,.
\end{eqnarray}

As $\Im q^2 b(\i q) \sim -q$ for large $|q|$, it is again
advantageous to split the integrals at some $q_0$ and, for
$q > q_0$, to rewrite the imaginary part appearing
in the integrand of \eqref{eqn:Z_force_num2}
to
\begin{eqnarray}
  \label{eqn:Z_force_num3}
  \Im\left[\e^{\i q
      (\eta_\parallel + 3 \mu^2/2)} \left(a_3(\i q)
    \e^{-\frac{3}{2} \mu^2  \left(q^2 b(\i q) + \i q\right)}
  - \e^{-\frac{\mu^2 q^2 + 3 \i q \mu^2}{2}}\right)\right]\nonumber\\
\end{eqnarray}
and the real part appearing in \eqref{eqn:deriv_Z_force_num} to
\begin{eqnarray}
  \Re\left(\e^{\i q (\eta_\parallel + 3 \mu^2/2)} a_3(\i q)
    \e^{-\frac{3}{2} \mu^2 \left(q^2 b(\i q) + \i
        q \right )}\right) \,.
\end{eqnarray}

In both cases the integrand is holomorphic for $\Im q < 0$. Hence the
integrals vanish if $\delta \eta_\parallel := \eta_\parallel +
3\mu^2/2 < 0$ which we already understood in the simple geometric
picture of the problem.

Both integrals now vanish in the limit of large $\eta_\parallel$ and
have well-behaved integrands on $[0, \infty]$. The precision with
which $\tilde f(\eta_\parallel, \mu)$ can be calculated is, however,
still limited by the relative error in evaluating the integrals. This
relative error grows quickly with increasing $\eta_\parallel$
limiting the range of $\eta_\parallel$ over which $\tilde
f(\eta_\parallel, \mu)$ can be calculated reliably (note that the
first term of \eqref{eqn:Z_force_num2} vanishes with increasing
$\eta_\parallel$ as well).

\end{appendix}


\begin{thebibliography}{99}
\bibitem{footnote:1} Actually one
  has to be slightly more precise and say that one considers a case
  where the obstacle is fixed and rigid on time scales where the
  thermal fluctuations of the filament equilibrate.
\end{thebibliography}
\end{document}